\documentclass{article}

\usepackage{arxiv}

\usepackage[utf8]{inputenc} % allow utf-8 input
\usepackage[T1]{fontenc}    % use 8-bit T1 fonts
\usepackage{hyperref}       % hyperlinks
\usepackage{url}            % simple URL typesetting
\usepackage{booktabs}       % professional-quality tables
\usepackage{amsfonts}       % blackboard math symbols
\usepackage{nicefrac}       % compact symbols for 1/2, etc.
\usepackage{microtype}      % microtypography
\usepackage{cleveref}       % smart cross-referencing
\usepackage{lipsum}         % Can be removed after putting your text content
\usepackage{graphicx}
\usepackage{natbib}
\usepackage{doi}
\usepackage{tikz}

\newcommand{\tikzcircle}[2][1pt]{%
  \tikz[baseline=(char.base)]%
    \node[draw,circle,inner sep=#1] (char) {#2};%
}

\title{Towards Next Generation Data Engineering  Pipelines}

\date{}

\newif\ifuniqueAffiliation
\uniqueAffiliationtrue

\ifuniqueAffiliation % Standard variant of author block
\author{ \href{https://orcid.org/0009-0003-7595-532X}{\includegraphics[scale=0.06]{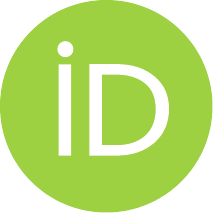}\hspace{1mm}Kevin M.~Kramer}\\
	University of Hagen\\
    Hagen, Germany\\
	\href{mailto:kevin.kramer@fernuni-hagen.de}{\texttt{kevin.kramer@fernuni-hagen.de}}\\
	\And
	\href{https://orcid.org/0000-0002-5960-5886}{\includegraphics[scale=0.06]{orcid.pdf}\hspace{1mm}Valerie~Restat} \\
	University of Hagen\\
    Hagen, Germany\\
	\href{mailto:valerie.restat@fernuni-hagen.de}{\texttt{valerie.restat@fernuni-hagen.de}}\\
	\AND
	\href{https://orcid.org/0009-0001-8848-1368}{\includegraphics[scale=0.06]{orcid.pdf}\hspace{1mm}Sebastian~Strasser} \\
	University of Regensburg\\
    Regensburg, Germany\\
	\href{mailto:sebastian.strasser@ur.de}{\texttt{sebastian.strasser@ur.de}}\\
	\And
	\href{https://orcid.org/0000-0003-2771-142X}{\includegraphics[scale=0.06]{orcid.pdf}\hspace{1mm}Uta~Störl} \\
	University of Hagen\\
    Hagen, Germany\\
	\href{mailto:uta.stoerl@fernuni-hagen.de}{\texttt{uta.stoerl@fernuni-hagen.de}}\\
    \And
	\href{https://orcid.org/0000-0003-0551-8389}{\includegraphics[scale=0.06]{orcid.pdf}\hspace{1mm}Meike~Klettke} \\
	University of Regensburg\\
    Regensburg, Germany\\
	\href{mailto:meike.klettke@ur.de}{\texttt{meike.klettke@ur.de}}\\
}

\else
\usepackage{authblk}

\newbox{\orcid}\sbox{\orcid}{\includegraphics[scale=0.06]{orcid.pdf}} 
\author[1]{%
	\href{https://orcid.org/0009-0003-7595-532X}{\usebox{\orcid}\hspace{1mm}Kevin M.~Kramer\thanks{Email address of the corresponding author: \texttt{kevin.kramer@fernuni-hagen.de}}}%
}
\author[1]{%
	\href{https://orcid.org/0000-0002-5960-5886}{\usebox{\orcid}\hspace{1mm}Valerie~Restat}%
}
\author[2]{%
	\href{https://orcid.org/0009-0001-8848-1368}{\usebox{\orcid}\hspace{1mm}Sebastian~Strasser}%
    }
\author[1]{%
	\href{https://orcid.org/0000-0003-2771-142X}{\usebox{\orcid}\hspace{1mm}Uta~Störl}%
}
\author[2]{%
	\href{https://orcid.org/0000-0003-0551-8389}{\usebox{\orcid}\hspace{1mm}Meike~Klettke}%
    }

\affil[1]{Faculty of Mathematics and Computer Science, University of Hagen, Hagen, Germany}
\affil[2]{Faculty of Computer Science and Data Science, University of Regensburg, Regensburg, Germany}
\fi

\renewcommand{\headeright}
\hypersetup{
pdftitle={Towards Next Generation Data Engineering  Pipelines},
pdfsubject={cd.db},
pdfauthor={Kevin M.~Kramer, Valerie~Restat, Sebastian~Strasser, Uta~Störl, Meike~Klettke},
pdfkeywords={data engineering pipelines, optimization, data quality, self-awareness, self-adaptation},
}

\begin{document}
\maketitle
\vspace{-1em}
\begin{abstract}
	Data engineering pipelines are a widespread way to provide high-quality data for all kinds of data science applications. However, numerous challenges still remain in the composition and operation of such pipelines. Data engineering pipelines do not always deliver high-quality data. By default, they are also not reactive to changes. When new data is coming in which deviates from prior data, the pipeline could crash or output undesired results. We therefore envision three levels of next generation data engineering pipelines: optimized data pipelines, self-aware data pipelines, and self-adapting data pipelines. Pipeline optimization addresses the composition of operators and their parametrization in order to achieve the highest possible data quality. Self-aware data engineering pipelines enable a continuous monitoring of its current state, notifying data engineers on significant changes. Self-adapting data engineering pipelines are then even able to automatically react to those changes. We propose approaches to achieve each of these levels.
\end{abstract}

% keywords can be removed

\section{Motivation}
In the last decades, the size of data which is collected and processed grew drastically and still continues to increase. While this huge amount of data offers great potential for gaining knowledge through analysis, many challenges remain in the handling and processing of it~\cite{4Generations}.
What makes the analysis especially difficult is poor data quality. Data quality refers to many different aspects such as availability, usability, reliability, relevance, and presentation quality~\cite{Cai2015}. In this work, we define high-quality data as data that contain as few errors as possible. Data acquisition usually does not follow a standardized process. Data can also be loaded from different sources and for different purposes. This leads to data having missing values, varying formats / schemas, duplicates, or other errors~\cite{GouDa}. Therefore, data cleaning is an important step in every data science project in order to yield uniform high-quality data~\cite{Ilyas2019}. Multiple sequential steps are required to create such a format from raw data. These steps are commonly executed in a so-called \textit{data engineering pipeline} which can be seen as an abstraction supporting the whole data engineering process from ingestion over data cleaning and transformation to storing the data in the desired format~\cite{4Generations,DERM}.  
A pipeline can be represented in different formats, e.g.\ in a graphical interface, as a script, or in a data format containing descriptions of the data engineering steps~\cite{Matskin2021ASO}. Generally speaking, data preparation, including collecting, cleaning, and organizing data, makes up 80\% of the work of a data scientist\footnote{CrowdFlower Data Science Report (2016) \url{https://visit.figure-eight.com/rs/416-ZBE-142/images/CrowdFlower_DataScienceReport_2016.pdf}}. This bears great potential for cost reduction, if data engineering processes for achieving high quality data are automatically done by an autonomous system.

\begin{figure}[ht]
    \centering
    \includegraphics[width=0.85\textwidth]{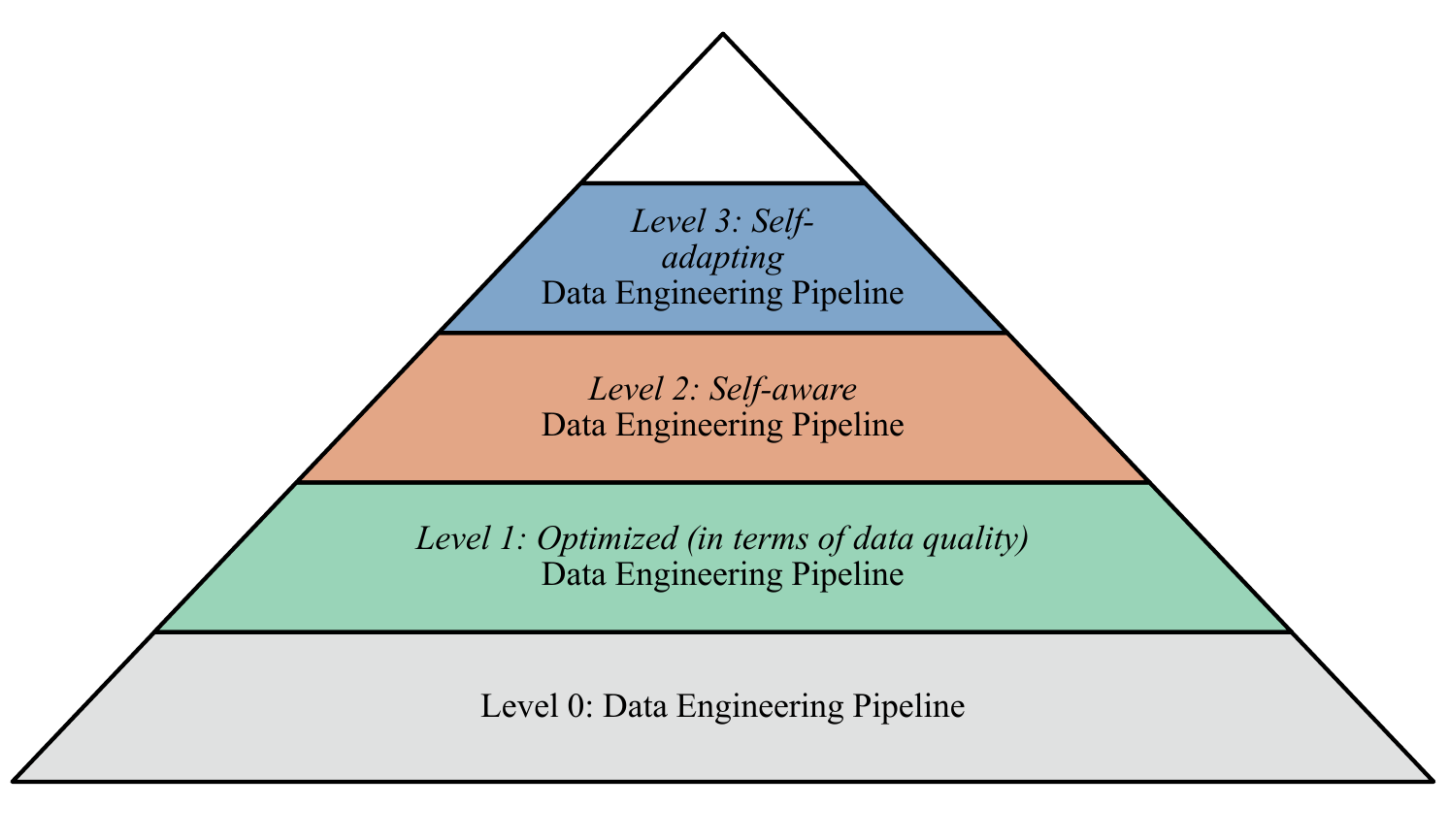}
	\caption{Different levels of next generation data engineering pipelines}
    \label{fig:pyramid_v2}
\end{figure}

On this basis, we envision three levels of next generation data engineering pipelines (see Figure~\ref{fig:pyramid_v2}). These go beyond the existing functionalities of data engineering pipelines. Once all levels are met, the result is an autonomous system that can be used to create, monitor, and adapt pipelines that lead to high data quality. For \texttt{\textit{Level 1}}, we require a data engineering pipeline to be \textit{optimized}. This means that the pipeline outputs (or intends to output) data with the best possible \textit{data quality}. 
\texttt{\textit{Level 2}} then focuses on the runtime phase of the data engineering pipeline. Here, we require a pipeline to detect changes in input and intermediate data and in the pipeline itself. This way, data and pipeline quality is secured not only at design phase, but throughout the whole runtime phase. 
Such pipelines can be called \textit{self-aware}. They enable a more targeted root cause analysis of errors for data engineers and provide more transparency of the entire process.  In \texttt{\textit{Level 3}}, we require data engineering pipelines to be \textit{self-adapting}. Like in related disciplines, self-adaptive workflows require self-awareness~\cite{Petrovska2021}. Once a significant change is detected by the system, the pipeline evaluates if it needs to be adapted and also triggers this adaptation if necessary. Thus, a self-adapting data engineering pipeline can be seen as an autonomous system which constantly observes its current state and changes itself if necessary in order to constantly provide a functional data engineering pipeline which delivers high quality data.

In the following, we will explain the different levels of data engineering pipelines. The focus is currently on structured and hierarchical data, arriving as a sequence of \textit{batches}. We assume each batch to have a \textit{schema}, but \textit{schema evolution} might occur over time, i.e.\ between batches. Other types of change over time are possible which are presented in the following. Also, there is uncertainty about the time, form, and extent of change regarding a new data batch coming from upstream.

For illustration purposes, we will look at an \textit{eye tracking use case} which is described in~\cite{et_jenny}. For our running example, we only consider a small portion of the data.
This data set contains the following properties\footnote{We use the term \textit{property} rather than e.g. attribute in order to refer not only to relational-like data. This follows the definition of property as a \emph{unifying term} in \cite[p.11]{Koupil2022}.}:
\begin{itemize}
    \item \textit{ET-GazeLeft-X}, \textit{ET-GazeLeft-Y}: The X and Y coordinates of the left eye of the participant
    \item \textit{ET-GazeRight-X},\textit{ ET-GazeRight-Y} The X and Y coordinates of the right eye of the participant
    \item \textit{Fixation-X}, \textit{Fixation-Y}: The X and Y coordinates of the participant’s fixation point on the screen 
    \item \textit{Group}: The study group to which the participant has been assigned
\end{itemize}

\begin{figure}[ht]
    \centering
\includegraphics[width=\textwidth]{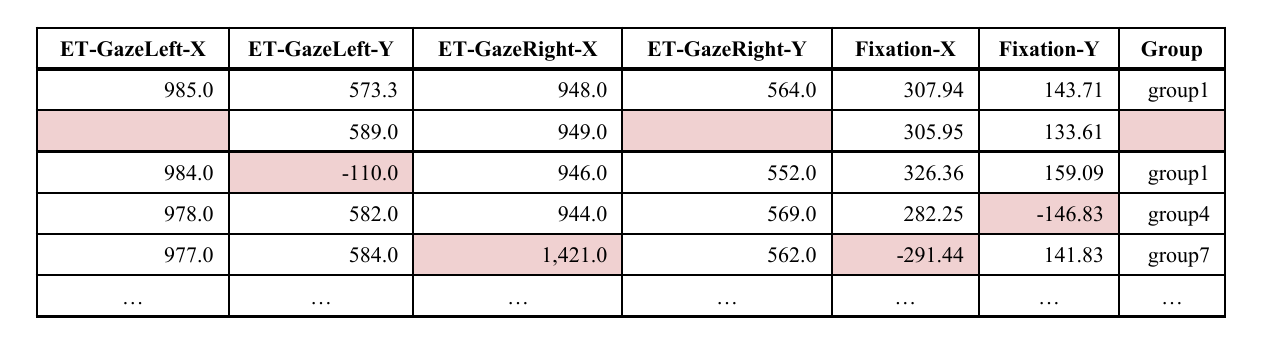}
	\caption{Exemplary selection of the eye tracking data, errors (missing values and interval violations) are shaded in red}
    \label{fig:et-dataset}
\end{figure}

An exemplary selection of the data can be seen in Figure~\ref{fig:et-dataset}. There are interval violations and missing values in the properties \textit{ET-GazeLeft-X}, \textit{ET-GazeLeft-Y}, \textit{ET-GazeRight-X}, and \textit{ET-GazeRight-Y}. \textit{Fixation-X} and \textit{Fixation-Y} also have interval violations. The property \textit{Group} contains missing values as well. The erroneous values are highlighted in red.

These data quality issues must be cleaned in order to ensure the quality of the analysis results. The first step is therefore to build an optimized data engineering pipeline in terms of data quality. It must thoroughly clean up the errors, thereby ensuring the highest possible data quality. 

\begin{figure}[ht]
    \centering
\includegraphics[width=\textwidth]{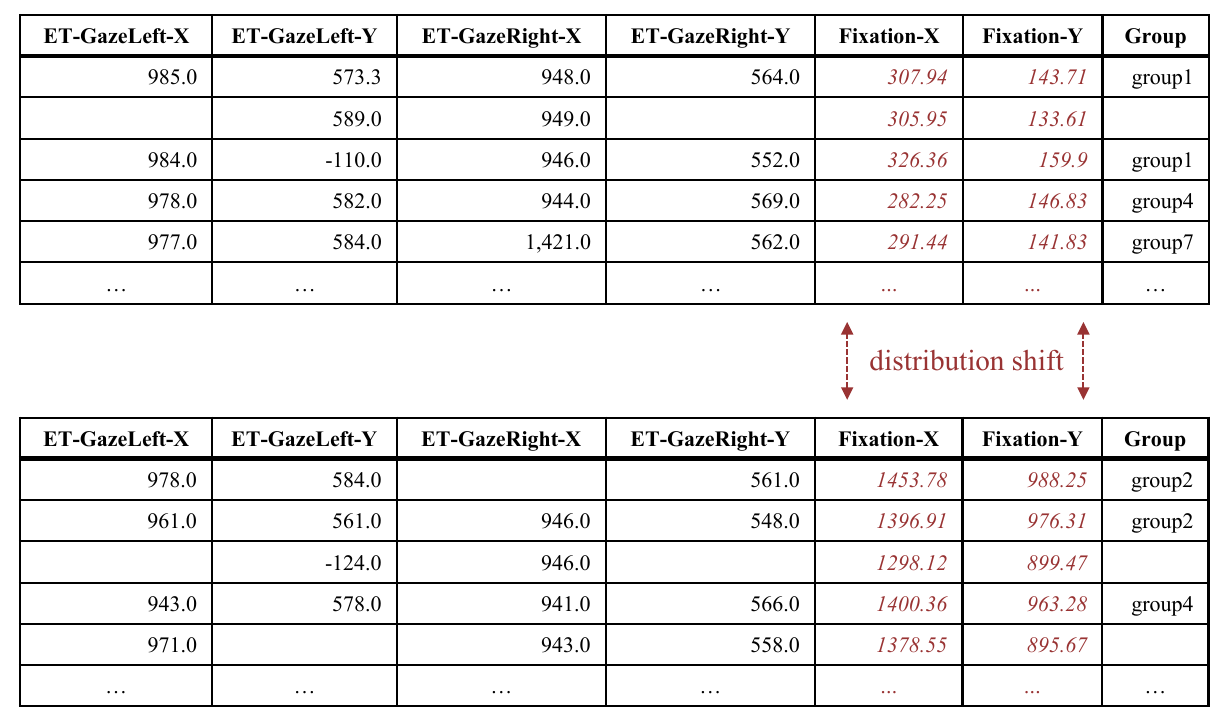}
	\caption{Semantic change: The distributions of the values of \textit{Fixation-X} and \textit{Fixation-Y} have changed}
    \label{fig:et-dataset-semantic-change}
\end{figure}

Once this pipeline has been found, it is deployed. Assume that, after some time in production, the screen for the eye tracking experiment is replaced. This changes the distribution of the values of \textit{Fixation-X} and \textit{Fixation-Y}. A \textit{semantic change} has therefore occurred. This is illustrated in Figure~\ref{fig:et-dataset-semantic-change}. The pipeline can now lead to incorrect results, since the interval limits have changed. For this reason, such a change must be detected automatically. In other words, the pipeline must be self-aware. If a significant change is detected through monitoring, the pipeline must also be automatically adapted so that it produces correct results again. This would correspond to a self-adapting pipeline. The recognition of changes in  the pipeline and its automatic adaptation leads to the best possible data quality again.

\begin{figure}[ht]
    \centering
\includegraphics[width=\textwidth]{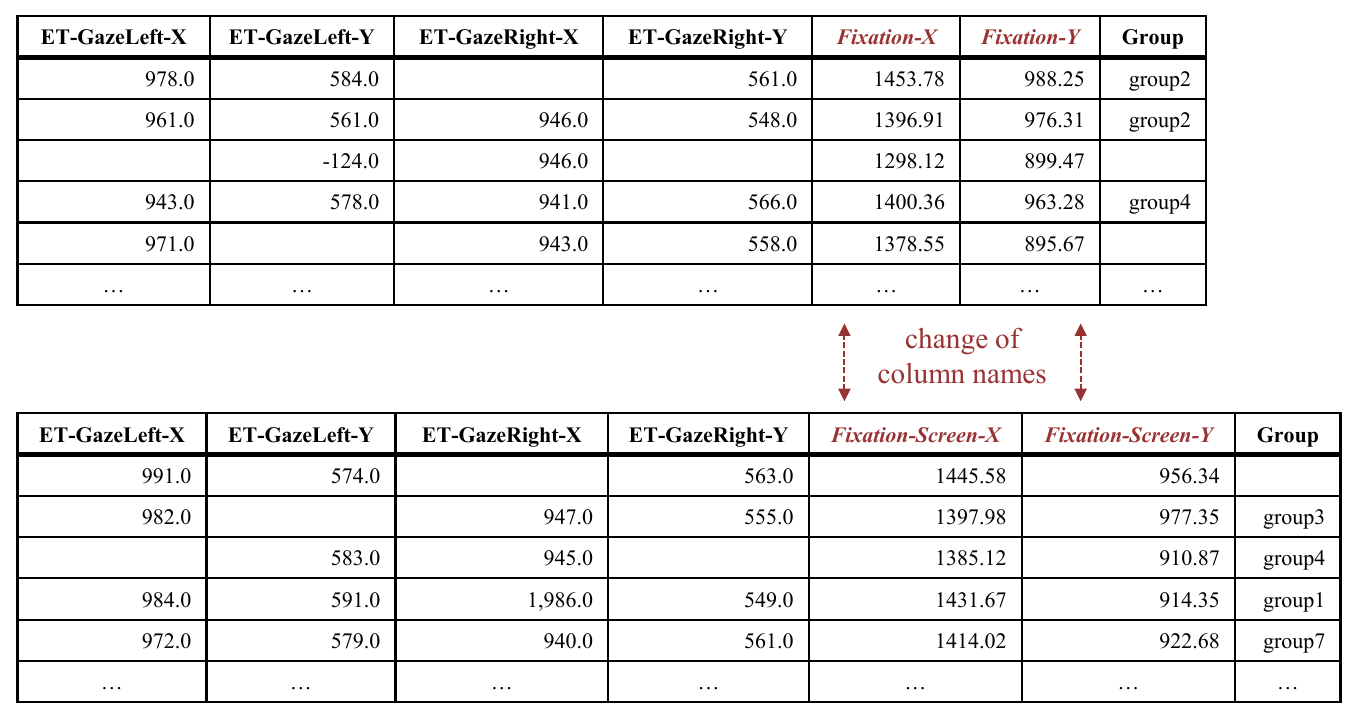}
	\caption{Structural change: The properties \textit{Fixation-X} and \textit{Fixation-Y} are now called \textit{Fixation-Screen-X} and \textit{Fixation-Screen-Y}}
    \label{fig:et-dataset-structural-change}
\end{figure}

Another possible scenario would be an update on the eye tracking device. As a result, the properties \textit{Fixation-X} and \textit{Fixation-Y} are now called \textit{Fixation-Screen-X} and \textit{Fixation-Screen-Y}, as can be seen in Figure~\ref{fig:et-dataset-structural-change}. A \textit{structural change} has occurred. This change would cause the pipeline to break. To prevent this, the self-awareness component of the pipeline performs automated inspection, senses this change and then triggers an adjustment by the self-adaptation subsystem. After the adaptation, the pipeline can continue to run and produce correct results. 

In the following sections, we first give an overview on the state of the art for the different levels we defined for data engineering pipelines in Section~\ref{sec:sota}. Afterwards, we propose approaches such that optimized (Section~\ref{sec:automatic_composition}), self-aware (Section~\ref{sec:inspection}), and self-adapting (Section~\ref{sec:evolution}) data engineering pipelines can be conceptualized and implemented. We also present an architecture and a technical outline of a system which enables building and operating these pipelines in Section~\ref{sec:system}. At last, we summarize the key points of our concept and identify research challenges we plan to tackle in the next steps in Section~\ref{sec:conclusion}.

\section{State of the Art}~\label{sec:sota}
Due to the huge relevance of data engineering in both academia and industry, plenty of research and tools exist in this field. In this section, we first introduce data engineering pipelines in general. Afterwards, we look at approaches and tools which lay the groundwork for achieving the proposed levels of data engineering pipelines.

\subsection{Level 0 : Data Engineering Pipeline}~\label{subsec:sota_general}
Data Engineering is a fundamental task in most scenarios dealing with data. Yet, it is often viewed as a means to an end, before the actual work can start. Therefore, the goal of this subsection is to give a broad overview of the theoretical and formal view on data engineering.
In an earlier work~\cite{4Generations}, we defined data engineering as "providing data for analysis", which includes pre-processing techniques for preparing data for further analysis in domains such as business intelligence and machine learning. In this work, we introduced different generations of data engineering approaches, starting with tasks like "Data Understanding and Data Profiling", "Cleaning and Data Correction", and "Data Transformation". 

Next, we classified chains of algorithms where each algorithm was developed to solve a specific data engineering problem. Following from this, we envisioned automatic composition of data engineering pipelines. The concept for this will be further developed here. 
The authors of~\cite{DERM} propose a reference model for data engineering which is based on a literature review and was verified empirically. They argue that such a model is necessary to successfully develop data engineering solutions, which consider the complete lifecycle of data. Their model is structured into phases which define specific tasks, such as create or transform. There are also layers which describe the context, for instance metadata or technology, in which the task is performed. In~\cite{DBLP:conf/dawak/RomeroW20} the authors assert that before any knowledge can be gained from data through analysis, pre-processing is mandatory, including data integration, data cleaning, duplicate elimination and storage. The authors argue that there is a need for a conceptual model and corresponding data engineering architecture. Key aspects of their envisioned model are a domain specific language, which is used to define the abstract, conceptual model of a pipeline, the use and management of metadata, pipeline execution optimization, and self-adaptive functionalities, e.g.\ automatically selecting the best data model for a given task. In the context of data science, the authors of~\cite{DBLP:conf/icse/BiswasWR22} found data engineering to be an integral part in all their study scenarios, including data science pipelines from a theoretical and practical viewpoint. Their findings suggest that the pre-processing step consists of data acquisition, data preparation, and storage and that these actions can be found in nearly all data science pipelines.

Different perspectives on data engineering exist in the literature. One way to look at it is through a specific domain. For example, the author of~\cite{AI-DE} puts a special focus on data engineering for AI systems and reviews the literature from this domain including architectures, technologies and its inherent challenges. Another perspective is to look at the kinds of data being processed. For instance, dealing with big data, i.e.\ data which fulfills the four Vs volume, velocity, variety, and veracity~\cite{BigData-4Vs}, in the context of data engineering bears its own challenges and potentials, as the authors of~\cite{BigData-DE} suggest. 

To manage the execution of data pipelines, data pipeline orchestration tools are often utilized, e.g., dagster\footnote{\url{https://github.com/dagster-io/dagster}} or prefect\footnote{\url{https://github.com/PrefectHQ/prefect}}. These tools allow to automate and monitor the complete data engineering pipeline starting with the ingestion of raw data and ending with the storage of the cleaned data. They provide extensive functionality for simplifying the process of moving data from source to target systems and monitoring data preparation processes. Take for instance a system which ingests large-scale data from multiple sources, cleans the data, and stores it in a data warehouse for further analysis. Here, a data pipeline orchestrator enables the data engineer to overlook all the steps in the pipeline. Alerts are triggered if a critical step of the pipeline fails. One example would be a process which performs data cleaning failing because there is a runtime error in the executed program. Besides these types of systems which usually arise from non-academic sources, scientific workflow systems offer similar functionality but are the product of research endeavors. The authors of~\cite{DBLP:journals/csur/LiewAGA0H17} present an overview of the latter and their workings, while in \cite{Matskin2021ASO} academic and non-academic systems are compared. 

\subsection{Level 1 : Optimized Data Engineering Pipeline}
\label{subsec:sota-level-1}
The intent to find the best possible pipeline also exists in the AutoML area. The aim is to create a pipeline that improves the efficiency of machine learning applications -- as automatically as possible. However, it has been shown that full automation cannot usually be achieved and that human involvement is necessary in many places~\cite{Karmaker2022}. A study with users from a wide range of domains has also shown that the focus of AutoML is mainly on automating model training. Data preparation is often neglected and has to be done with other tools~\cite{Xin2021}.

Many different tools and methods for data cleaning and thus improving data quality exist (e.g.~\cite{boostclean, holoclean, baran, katara, Hameed2020}). Nevertheless, these approaches only focus on individual error types or require a great deal of manual effort to find the best pipeline. An end-to-end solution that generates an optimized pipeline with regard to data quality is still lacking~\cite{Abedjan2016}.

A framework for the optimization of data cleaning pipelines named SAGA is presented in~\cite{Siddiqi2023}. Ideas from AutoML are adapted. It is also fully integrated into Apache SystemDS~\cite{Boehm2020}. SAGA covers outlier handling, missing value imputation as well as the handling of class imbalance and / or flipped labels. Others primitives can be added. The evolutionary algorithm is utilized to find a set of candidate logical pipelines. These then serve as a starting point for physical pipeline tuning. The paper also discusses various strategies for parallelization. This is also interesting for our vision.  However, the focus of this work is specifically on data cleaning for machine learning, which has unique characteristics~\cite{Siddiqi2023}. In addition, as described, only some of the possible algorithm classes for data cleaning (like missing value imputation and handling of violated functional dependencies) have been taken into account so far.

The optimization of data cleaning pipelines is also the objective in~\cite{Neutatz2021}. Optimizers are used to combine different data cleaning information (here called "signals") to produce a data cleaning pipeline. Anyhow, the degree to which different signals and optimizers can be combined in a holistic data cleaning approach remains an open research question. Moreover, the scope here is cleaning for machine learning as well. In our vision, we aim for a more general solution that can also be used independently of the downstream application.

In order to achieve such a solution, in \texttt{\textit{Level 1}} we optimize with regard to data quality. For this reason, an important aspect is the evaluation of data quality.
In this context, Sambasivan et al.~\cite{Sambasivan2021} emphasize the importance of \textit{goodness-of-data}. They argue that there needs to be a shift from goodness-of-fit, as measured by well-established metrics such as F1, Accuracy and AUC, to goodness-of-data. They also note that there is no standardized metric for goodness-of-data and emphasize the need for further research in this area. Even though data quality has been researched for more than 30 years~\cite{Wang1993, Wang1995} and various metrics exist (e.g.~\cite{Blake2011, Heinrich2018}), they only cover part of data quality. A system named Deequ for evaluating data quality is described in~\cite{Schelter2018}. They take completeness, consistency, and accuracy into account. Different evaluations are possible through the combination of quality constraints and custom validation code. For example ``isComplete'' would be a constraint to check for missing values in a column and ``hasSize'' a custom validation for the number of records. The authors of~\cite{Elouataoui2022} look at data quality specifically in the context of big data. For this purpose, they present a framework consisting of 12 metrics. There are also different data verification methods from research and practice, which are automated to varying degrees. However, none of them covers all possible error types. Prominent examples for open-source tools are pandera\footnote{\url{https://github.com/unionai-oss/pandera}}, Great-expectations\footnote{\url{https://github.com/great-expectations/great_expectations}}, and pydantic\footnote{\url{https://github.com/pydantic/pydantic}}. Different data verification methods also exist in research. Systems such as HoloDetect~\cite{holodetect} or Raha~\cite{raha} aim to detect data errors with as little manual effort as possible. They utilize machine learning for this purpose. Other approaches such as~\cite{Shrestha2023} and~\cite{Bors2018}, on the other hand, focus on manual evaluation with the involvement of domain experts. With CheDDaR (Checking Data -- Data Quality Review)~\cite{CheDDaR, CheDDaR2025}, we have proposed a framework for the evaluation of data quality in previous work. This can be used flexibly, depending on how much domain knowledge is available or whether ground truth exists. Overall, however, there is still no standardized solution or single metric for evaluating data quality.

\subsection{Level 2 : Self-aware Data Engineering Pipeline}
Insights of several research directions can be utilized in order to implement systems which enables data engineering pipelines to be self-aware. Significant effort has been put into the debugging of data pipelines for data science applications. Several tools (e.g.~\cite{Murta2014, Chapman2020}) exist which capture provenance information from data preparation pipelines. This provenance information is used to depict how different operators change the data and therefore help users in debugging data pipelines. Other approaches are specialized for data preparation pipelines for machine learning models and look into machine learning-specific information like the representation of groups for fairness measurement~\cite{Grafberger2022, Biswas2021} or the estimated impact of a pre-processing operator on model outcome~\cite{Zelaya2019}. In~\cite{Epperson2024}, the authors suggest an automated continuous data profiling throughout the data science pipeline. This is meant to guide the user in finding potential issues in their pipeline. While most of the aforementioned approaches and tools aim at finding issues of pipelines during the design phase, we see lots of potential in incorporating numerous concepts in our self-awareness component as we also aim at detecting issues in the data engineering pipeline. The collection of provenance information can help in getting an overview of how the data is processed throughout the pipeline.
Continuous data profiling, i.e., the collection of data about data (= metadata)~\cite{Abedjan2019}, is a crucial part of our concept, too. An example would be if the same data profiling procedure yields highly different results on consecutive input batches. This could indicate a change in the data source which in turn would necessitate a change in the pipeline.

Another field of research which is relevant in the context of self-aware data engineering pipelines is data validation. Validating input data is an important part of our concept as data with changing or undesired characteristics can produce errors or unexpected behavior in pipeline operators. Defining and checking against user-defined constraints is an approach which is often used in this context. We introduced Great-expectations and Deequ~\cite{Schelter2018} as tools for evaluating data quality in Section~\ref{subsec:sota-level-1}. They can be also be used for validating new incoming data. Deequ does this by utilizing common data quality metrics and user-defined constraints. It also contains basic approaches for suggesting constraints. This idea is enhanced in other work~\cite{Heine2019} where more complex data quality rules can be automatically detected in data. Other data validation approaches compare fresh data with historical data based on data summaries which contain descriptive statistics over the columns and generate alerts when there is a deviation between these summaries (based on a certain threshold)~\cite{Tu2023, Shankar2023}. In our work, we plan to develop a hybrid approach where we semi-automatically define constraints based on historical data and data summaries. An important requirement for these constraints is that they are easily adaptable in the case of changing input data or environment. 

We are aiming for a system which continuously monitors data engineering pipelines and detects changes in input and operator behavior reliably. One part of this is monitoring the quality of the data which is flowing through the pipeline. The authors of~\cite{Ehrlinger2019} and~\cite{Lettner2020} propose the system DaQL which continuously monitors the data quality in pipelines providing data for machine learning models. The authors of~\cite{Narayanan2024} extend this concept of monitoring onto the whole software application which implements the data pipeline, thus providing observability on the complete data engineering pipeline. We envision a general-purpose system for providing observability into data engineering pipelines which can be configured easily based on the concrete application.

\subsection{Level 3 : Self-adapting Data Engineering Pipeline}
\label{subsec:sota-level-3}
Software systems change constantly over time. This often leads to new requirements on schemas. Multiple studies have been conducted that show schema evolution in real world scenarios~\cite{DBLP:conf/eScience/SchulerSVWBK23, DBLP:conf/er/ScherzingerS20, DBLP:conf/iceis/CurinoMTZ08}. Systems which autonomously deal with such changes are often described as having self-* properties, where the star implies a wildcard for a whole range of different aspects, including self-healing, self-maintenance, and so on. Besides giving an overview and definitions on self-* software, including self-awareness and self-adaptation\footnote{\textit{self-adaption} is a less frequently used term for \textit{self-adaptation}}, the authors of~\cite{Frei2013SelfhealingAS} propose a definition for robustness, namely "A system does not easily get disturbed in its normal functioning. It can cope with failures, changing conditions, and is able to remain usable". Even though this definition is rather lenient, it proposes a useful starting point. Achieving such a robust system state is easier when dealing with purposeful change. This type of change is not only known to the user but actively pursued. For example, there exist systems for deliberate co-evolution of data schema and application~\cite{DBLP:conf/icde/ScherzingerMK21, DBLP:conf/eScience/SchulerK22}. In our use-case this is not possible, since there is uncertainty about the time and extent of change. Dealing with uncertainty in the context of self-adaptive systems has been an active field of research~\cite{uncertainty1, uncertainty2}. Resolving uncertainty with respect to the time of change, i.e.\ whether or not significant change happened between data batches, is part of the self-awareness capabilities, as presented in Section~\ref{sec:inspection}. Interpreting the extent of change is the responsibility of the self-adaptation system, since this is the basis for any form of adjustment. This interpretation can be broken down into different categories concerning different areas. For instance, dealing with a changed schema and handling ambiguities regarding schema modification operations, as we proposed in \cite{DBLP:conf/bigdataconf/KlettkeAS0S17}, is an important requirement for adaptation. The authors of~\cite{DBLP:conf/sigmod/SpothKLHL21} propose a solution to automatically resolve schema ambiguities using heuristics in the context of JSON schema discovery. A recent work suggests to use large language models to resolve such schema ambiguities automatically~\cite{fu2024compoundschemaregistry}. 

Metadata with respect to operators, e.g.\ configuration parameters, or the data itself, e.g.\ the schema or data characteristics~\cite{Abedjan2019}, are often stored separately from the pipeline definition which can take several forms in itself~\cite{Matskin2021ASO}. This allows for flexible adjustments during runtime, since these files can be loaded dynamically. Apache Avro\footnote{\url{https://github.com/apache/avro}} is a data serialization framework that stores schema information in designated files and supports parts of schema evolution inherently. 

Different concepts and techniques for dynamically adapting software exist. In \cite{DBLP:journals/tse/ShevtsovBWM18}, the authors review the literature on control-theoretical software adaptation. LLMs have also been proposed for automatically improving software which can be seen as a general solution applicable to adaptation~\cite{DBLP:conf/ease/Ruiz24}. Dealing with uncertainty in the context of self-adaptation was investigated in~\cite{DBLP:journals/jcst/YangLTMXS13} where fuzzy logic was utilized.

We plan on combining these diverse functionalities into a coherent system which enables self-adaptation. In an earlier work we proposed conceptual requirements for self-adaptation capabilities \cite{DBLP:conf/gvd/Kramer23}. Even though it might seem that the individual components of said system already exist, selecting the right combination while assuring functionality and correctness is a major challenge which needs to be solved for our envisioned self-adapting data engineering pipeline.

\section{Next Generation Data Engineering Pipelines}
\begin{figure}[ht]
    \centering
\includegraphics[width=0.85\textwidth]{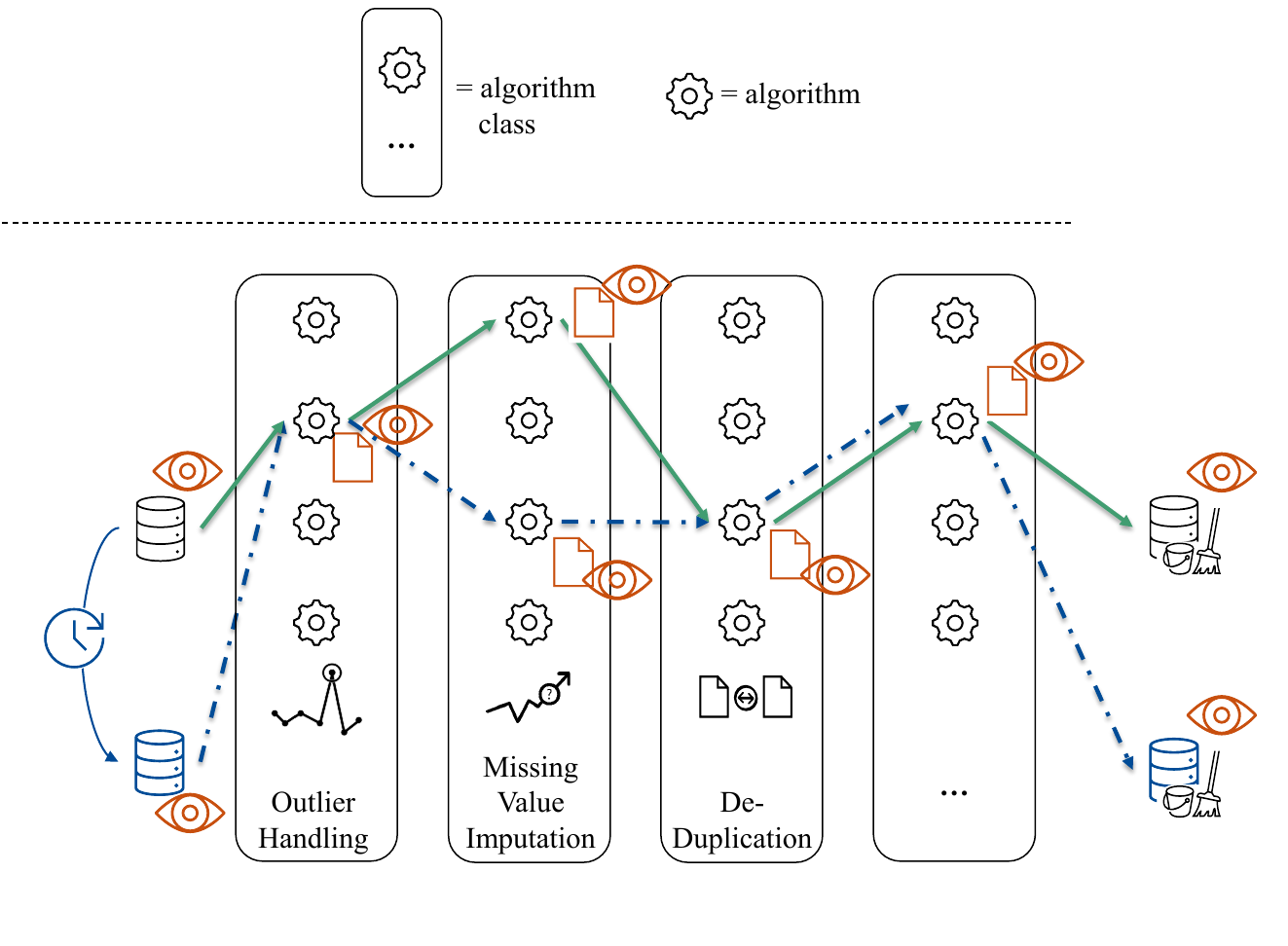}
	\caption{Overview of the proposed vision}
    \label{fig:overview}
\end{figure}

Our vision, which covers all levels of the pyramid, is shown in Figure~\ref{fig:overview}.

\textit{Level 0 (black boxes):}
A functioning data engineering pipeline consists of several steps. This corresponds to the base (grey) level of the pyramid. Different \textit{algorithms} from various \textit{algorithm classes} are available for the steps of a pipeline. In the eye tracking use case, for example, missing values need to be imputed and interval violations need to be corrected. The corresponding algorithm classes for this would be missing value imputation and the handling of interval violations. An algorithm is a concrete method, e.g.\ missing value imputation by imputing the mean value. In the overview in Figure~\ref{fig:overview}, algorithm classes are represented by slices and specific algorithms by gears. Instead of the terms \textit{algorithm} and \textit{algorithm class}, \textit{operator} / \textit{operator class} or \textit{method} / \textit{method class} can also be used.

\textit{Level 1 (solid, green arrows):}
In \texttt{\textit{Level 1}}, the goal is to build an optimized data engineering pipeline based on the initial data batch and requirements. In this context, optimized means a pipeline that leads to the best possible data quality. This corresponds to the first (green) level of the pyramid.

\textit{Level 2 (orange eyes):}
Once the optimized pipeline has been found and deployed, it must continuously monitor itself, thereby becoming self-aware. On the one hand, this concerns the changes made by the individual operators in the pipeline. On the other hand, changes between different data batches must also be monitored. This corresponds to the second (orange) level of the pyramid.

\textit{Level 3 (blue, dashed arrows):}
However, an optimized and monitored pipeline may break due to structural or semantic differences inherent to a new data batch. Therefore, the self-adapting pipeline must react to such change by evaluating the situation and adapting itself accordingly. The result is an altered pipeline which optimizes data quality for the new circumstances. This corresponds to the third (blue) level of the pyramid.

These three levels are described in more detail in the following subsections. For the sake of simplicity, we initially only differentiate here between the following data types: \textit{numerical}, \textit{categorical}, \textit{textual}.

\subsection{Level 1 : Optimized Data Engineering Pipeline}~\label{sec:automatic_composition}
\begin{figure}[ht]
    \centering
\includegraphics[width=0.85\textwidth]{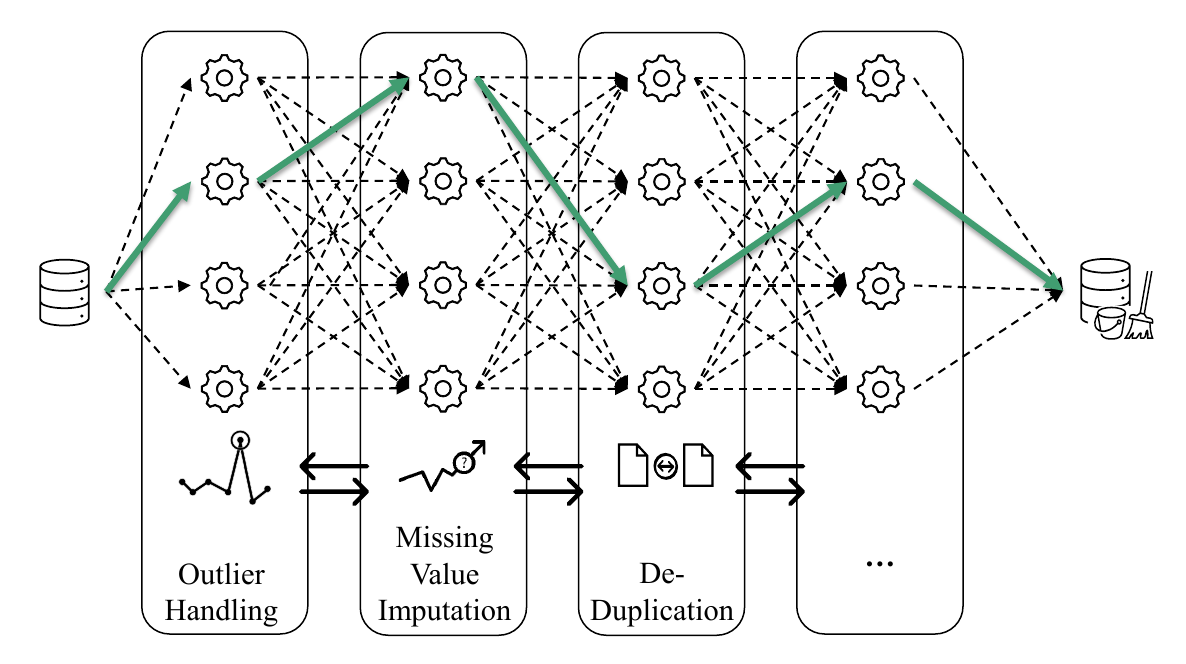}
	\caption{Automatic composition of an optimized data engineering pipeline. Optimized refers to the pipeline resulting in the best possible data quality (in this case represented by green, solid arrows).}
    \label{fig:compposition}
\end{figure}

As described in Section~\ref{subsec:sota-level-1}, although there are many individual data cleaning tools, there is no end-to-end solution that can be used independently of the downstream application. For this reason, the first level of our vision is the automatic composition of an optimized data engineering pipeline. As already described, such a pipeline consists of many different steps. Examples are outlier handling, missing value imputation, and deduplication. Many different algorithms exist for each of these steps. This creates a huge search space from which the best pipeline must be selected, as can be seen in Figure~\ref{fig:compposition}. The dashed lines indicate the search space. The solid green line represents the best pipeline. Since an operator is required for each error type in a property (for which there are different possibilities) and these can theoretically be placed in any order, the search space theoretically consists of $\sum_{i=0}^{N}x_i!$ possibilities, where $x_i$ is an operator and $N$ corresponds to the number of all operators in the pipeline. As described, for every $x_i$ itself, there are several possibilities. Of course, it must be taken into account here that not every operator and every order is useful for the specific use case. In addition, the order may not play a role for some operators, as they can be executed in parallel. Nevertheless, without in-depth domain knowledge, a large search space is created that cannot be searched manually.

As an example, consider the eye tracking use case. As described, the following error types can be found in the following properties: 
\begin{itemize}
    \item interval violations in \textit{ET-GazeLeft-X}, \textit{ET-GazeLeft-Y}, \textit{ET-GazeRight-X}, \textit{ET-GazeRight-Y}, \textit{Fixation-X}, and \textit{Fixation-Y}
    \item missing values in \textit{ET-GazeLeft-X}, \textit{ET-GazeLeft-Y}, \textit{ET-GazeRight-X} \textit{ET-Gaze\-Right-Y}, and \textit{Group}
\end{itemize}
In total, there are interval violations in six and missing values in five properties. Consequently, $5 + 6 = 11$ operators are required. This theoretically results in $11! = 39,916,800$ different possibilities of ordering the operators in a pipeline. As mentioned, it must be taken into account that operators could be independent of each other and therefore irrelevant for the order of the pipeline. Moreover, an operator may be used to repair the same error for similar properties. Nevertheless, it shows that the initial search space can be very big. In addition, there are many different concrete operators for each operator class. Even if we assume that there are only 10 different operators, the search space would already consist of $399,168,000$ different pipelines. If one also takes into account the fact that different parameters can be selected for most operators, it becomes clear that the search space can become extremely large. This highlights the need for automated optimization to find the best pipeline.

What exactly the "best" pipeline is also depends on the analysis objective. For example, if the data are used to train a machine learning model to predict customer behavior, the best pipeline is the one that delivers the best predictions in combination with the model. However, it is often the case that different teams work on data preparation and subsequent analysis. The team that is responsible for data preparation has no knowledge of the analysis objectives or how it is carried out. After preparation, the data is stored in a data lake, for example. From here, the data analysts can access it. The analysts in turn cannot influence data preparation. Moreover, different teams of analysts often access the data with different analysis objectives and methods~\cite{Krishnan2016}.

This should not contradict the emerging field of Data-Centric AI. Here, the aim is also to produce high-quality data sets~\cite{Zaharia2021}. In this context, as mentioned in Section~\ref{subsec:sota-level-1}, Sambasivan et al.~\cite{Sambasivan2021} emphasize the importance of goodness-of-data and the need for further research to find a standardized metric for this purpose. This fact, as well as the aforementioned practice of having different teams working on data pre-processing and different analysis objectives, underlines the need for a more general solution, as we describe in our vision of the optimized data engineering pipeline.

For this reason, a metric is required that can be used to determine the "best" pipeline independently of the subsequent analysis. The most important goal of data preparation is to deliver consistent, error-free data and to ensure the quality of the data~\cite{Johnson2024}. Thus, the quality of the pipeline is directly related to the quality of the data. The better the data quality, the better the pipeline. An optimized pipeline in this scenario thus refers to the pipeline resulting in the best possible data quality.

Even if access to the downstream analysis is not possible as described, we envision that constraints can be specified depending on the use case. This can for example entail that a certain property may not be imputed or deleted because the information is needed for fraud detection. Different use cases can introduce different constraints, which leads to not just one but several pipelines being created. 

\begin{figure}[ht]
    \centering
    \includegraphics[width=0.75\textwidth]{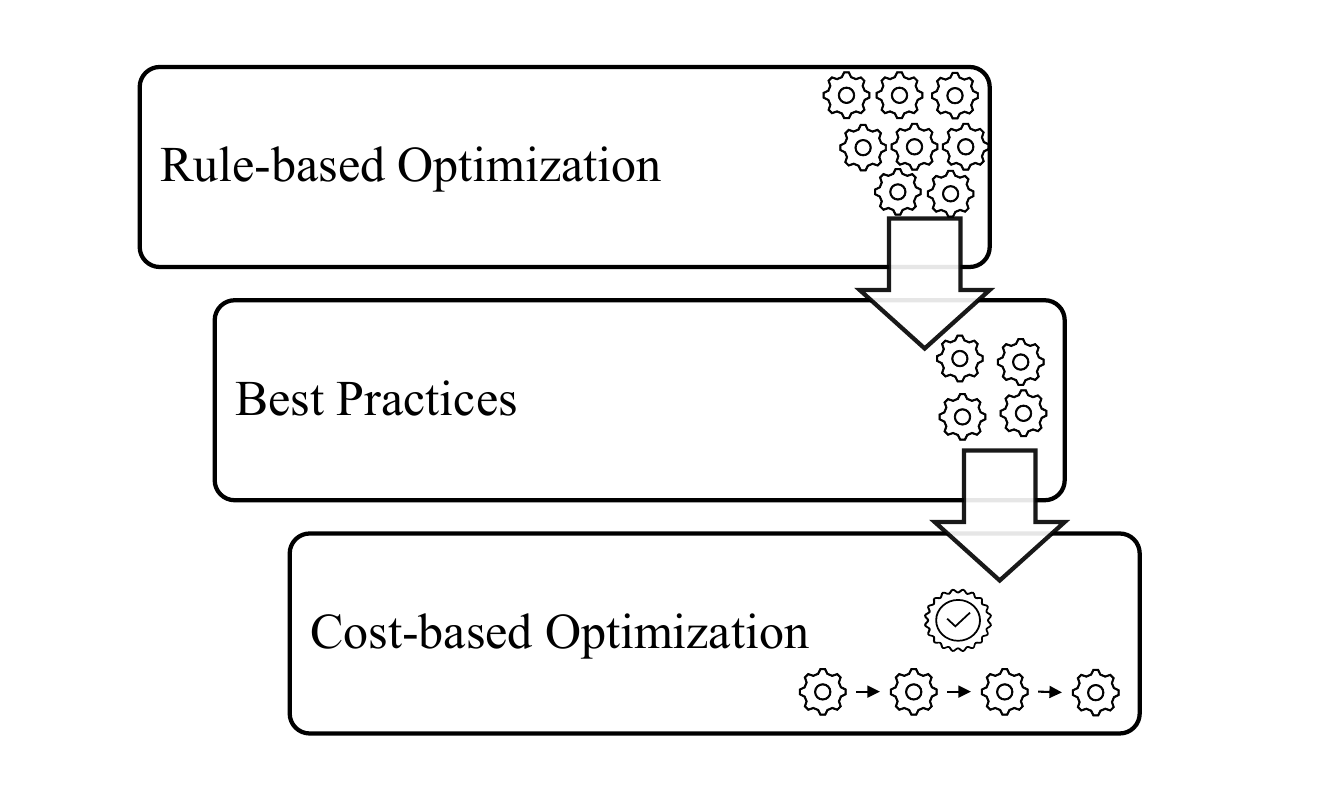}
	\caption{Overview of the process for determining an optimized data engineering pipeline}
    \label{fig:level1-ablauf}
\end{figure}

This subsection describes how the quality of a pipeline can be measured and how an automatic composition of one pipeline (or multiple pipelines) can be generated. The process is based on proven methods for database query optimization. An overview is shown in Figure~\ref{fig:level1-ablauf}. In~\cite{dataplat}, we introduced the idea of an end-to-end data quality optimizer. The procedure is outlined below and illustrated by the eye tracking example. A detailed description of the individual steps can be found in~\cite{dataplat}. In the following, we will outline the steps using our running example.

\subsubsection{Rule-based Optimization}
In order to build a suitable pipeline, it is first necessary to determine which errors need to be corrected. For this reason, the first step is \textit{error detection}~\cite{Abedjan2016}. An \textit{error profile} is created for this purpose. This is a JSON document that records all errors in the data set. Both the error type (e.g. \textit{missing value}) and the position of the error in the data set are documented. The appropriate algorithms for \textit{repairing} the errors can then be chosen from the available search space. As can be seen in Figure~\ref{fig:compposition}, there is a very large search space (dashed lines) from which the best pipeline (green, solid line) must be determined. As described, the quality of the pipeline is measured by the data quality of the cleaned data. First, the search space consisting of all algorithms has to be narrowed down. This is done using \textit{rule-based optimization}. Constraints are used to remove those operators from the search space that are unsuitable for the underlying data set. For example, if a textual or categorical property is given, a missing value imputation by mean imputation is not suitably applicable. This algorithm can only be used for numerical properties. Applied to the example of eye tracking data: As described, the \textit{group} property contains missing values. A missing value imputation must therefore be performed. The property only contains categorical values that specify the study group to which the participant was assigned (e.g.\ \textit{group1}, \textit{group2}, \textit{...}). A mean imputation is therefore not applicable in this case. Another example would be a case where the missing rate of a property is high. It was shown, that in this scenario the K-nearest Neighbors method leads to significantly worse results compared to other algorithms~\cite{Tsai22}.

\begin{figure}
    \centering
    \includegraphics[width=0.35\textwidth]{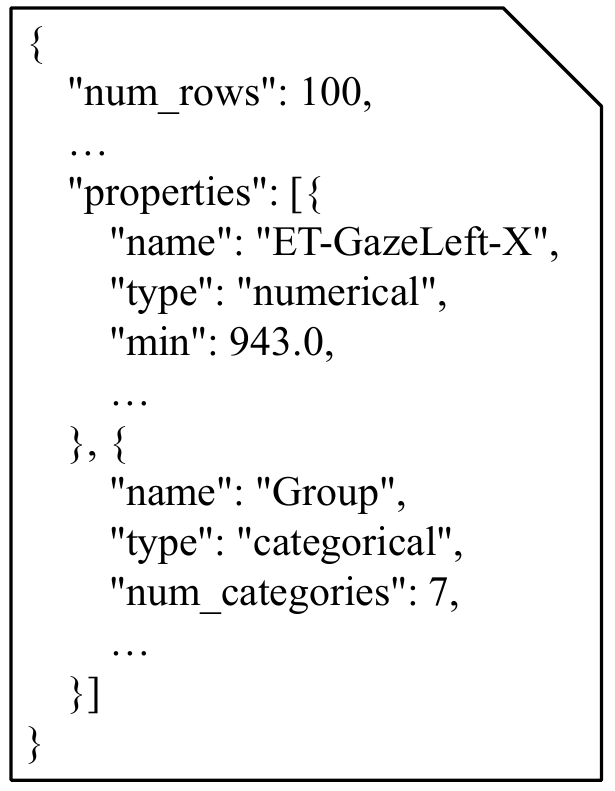}
    \caption{Data profile for the input dataset of our running example}
    \label{fig:data_profile}
\end{figure}

As a further development of the idea presented in~\cite{dataplat}, the aspect of \textit{data profiles} is addressed in more detail here. Metadata are required to find those operators that are not suitable for the data. Such metadata include, for instance, the data type, as shown in the previous example. Other important metadata concern the distribution of a property and the schema of the data itself. A lot of different types of metadata can be extracted from a data set. In order to cover different use cases and different data models, we plan to offer a highly configurable data profiling component. In order to secure comparability between different pipeline variants, a uniform representation for a data set with a concrete set of metadata is needed. We call this concrete set of metadata to describe a data set a data profile. Data profiles are important sources of information about the data~\cite{Abedjan2019}. In the case of our running example (see Figure~\ref{fig:et-dataset}), a data profile is shown in Figure~\ref{fig:data_profile}. As data profiling is a key ingredient for the downstream self-awareness component, we discuss the generation and the properties of a data profile in more detail in Section~\ref{sec:inspection}.

In the current step, the data profile is used to check which operators are not suitable for the data and thus reduce the search space. Several open research questions arise: First, such constraints must be found. An example of a constraint would be \textit{algorithm: mean value imputation} $\rightarrow$ \textit{permitted data types: numerical}. Once such constraints have been found and formalized, they must then be efficiently linked to the data profiles in order to obtain the relevant algorithms. To find such constraints and efficiently link it to the data profiles presents an open research question.

As a result of the first part of rule-based optimization, the search space only consists of those algorithms that are suitable for the data. In the next step, the search space can be further reduced by identifying the operators that are irrelevant regarding the order of the pipeline. Although these operators remain part of the pipeline, they can temporarily be removed from the search space because they can be executed at any time. Consider the following example of eye tracking data: Suppose we only considered the \textit{ET-GazeLeft-X} and \textit{ET-GazeLeft-Y} properties. If we replace the interval violations with the upper interval limit and the missing values with a fixed value within the permitted interval, the order of these two operators would be independent. Both variants would lead to the same result. The situation is different if we use the mean imputation. The calculation of the mean would lead to different results depending on whether we adjust the interval violations before or after the imputation. Even if it is very likely that only a few algorithms are independent of each other, this should still be taken into account, as any reduction of the search space is beneficial.

\subsubsection{Best Practices}
After this step, the search space only contains those algorithms that are suitable for the data and play a role in the order of the pipeline. To reduce the search space even further, \textit{best practices} can be applied in the next step. Best practices can be highly dependent on the use case and can be specified by domain experts. This allows them to contribute their knowledge. For example, if their experience has shown that a specific algorithm delivers particularly good results with the underlying data. In contrast to rule-based optimization, certain operators are not removed because they are not suitable, but algorithms are explicitly included in the pipeline (whereby other algorithms of the same algorithm class are removed from the search space).

In addition, the order of the pipeline or part of the pipeline can be explicitly defined here. This step enables an optional human-in-the-loop approach in the otherwise automated procedure. Even if this step is optional, it is very important. Any reduction in the search space can significantly improve the efficiency of \textit{cost-based optimization}. 

Regardless of the specific use case, a formalization of general best practices in form of \textit{heuristics} would also be beneficial. For example, Hasan et al.~\cite{Hasan2021} state that it is recommended for missing value imputation to standardize the data first when using distance-based algorithms (such as missing value imputation with K-nearest Neighbors). Finding further such heuristics remains an open research question.

\subsubsection{Cost-based Optimization}
Once the search space has been narrowed down as far as possible, the best pipeline must be selected from the remaining operators. This is done in the \textit{cost-based optimization}. In this step, the pipeline that leads to the best possible data quality is selected from many different possible pipelines. The data quality of the individual pipelines must thus be measured. A suitable metric is still an open research question. In~\cite{CheDDaR}, we describe a proposal for measuring data quality.
For this purpose, the different error types are divided into \textit{levels} -- depending on which part of the data set the error affects and how it can be measured. For example, a missing value can be detected simply by inspecting a single field of a property. The proportion of missing values can then be calculated as a percentage of all missing values of the corresponding property. The individual percentages must then be combined to form a collected metric.
In addition, the metric can be adapted to a use case using weights if required, e.g. for scenarios in which it is particularly important that there are no missing values in the data, but outliers can be neglected. Different variants are available for detecting errors -- depending on how much domain knowledge or reference data is available.

In addition to data quality, further optimization goals can be specified, e.g.\ for applications with limited resources or those in which the data must be processed near real time. The selection of a suitable algorithm for optimization is an open research challenge.

\subsubsection{Research Challenges}
\label{sec:auto_comp:reschal}
As described, \texttt{\textit{Level 1}}, the optimized data engineering pipeline, is subject to a number of research challenges. For one, constraints such as \textit{algorithm: mean value imputation} $\rightarrow$ \textit{permitted data types: numerical} must be identified. Then an efficient way must be found to link these constraints with the data profiles.

Another research challenge is analyzing which algorithms are independent of each other and therefore do not play a role in the order of the pipeline. Even if these are still relevant for the final pipeline, they lead to fewer permutations and hence to a smaller search space.

The reduction of the search space poses a further research challenge. General best practices in form of heuristics must be found. In addition to the selection of individual operators, a specific order of operators can also be defined here. This can lead to a significant shortening of the search space, which can considerably improve the efficiency of cost-based optimization.

In cost-based optimization, one challenge is the optimization in terms of data quality. For this purpose, the quality of the data must be measurable using a metric. Such a metric does not yet exist. In~\cite{CheDDaR} we present an initial proposal how such a metric could be obtained. Another research challenge is to select the suitable algorithm for the optimization. The search space can vary in size and depends on the dimensionality of the data. As described, the search space can become extremely large even with a small data set such as in the eye tracking example. The more properties there are in a data set, the more permutations are possible and, accordingly, the search space becomes even larger. The additional optimization goals -- besides data quality -- bring a further variation. Examples for these secondary optimization goals could be execution time or energy consumption.

\bigskip

The result of this process is the data engineering pipeline that leads to the best possible data quality. As this should be independent of the specific technology in which it is implemented, the result is created in the form of a \textit{pipeline profile}~\cite{ALPINE}. This is an abstract description of the pipeline in JSON. In addition to general characteristics such as the associated data set, it contains all information about the necessary operators and their parameters as well as the sequence in which the operators are executed. The implementation is carried out via \textit{adapters}, e.g.\ a script that implements the pipeline operators in Python. With these adapters, the selected pipeline is then deployed in production. Here, the pipeline must be monitored and potential changes detected automatically. This will be discussed in more detail in the next section.

\subsection{Level 2 : Self-aware Data Engineering Pipeline}
\label{sec:inspection}
In Section~\ref{subsec:sota_general}, data pipeline orchestrators were highlighted as tools which monitor data engineering processes and give feedback if components of the pipeline are not working properly, e.g.\ if there is a runtime error. However, they do not closely observe the data which is processed in production. For the initial composition of a pipeline as described in Section~\ref{sec:automatic_composition}, a fixed batch of data is used in order to find the optimal pipeline regarding data quality. But there is no guarantee that the pipeline is still optimal if unseen data runs through. For instance, new data can have a different schema than expected. This could lead to pipeline crashes in cases where the operators are not \textit{robust} to those changes in schema. The concept of robustness and the handling of different failure types in the pipeline will be explained more extensively in Section~\ref{sec:evolution}. In our running example, assume that an optimal pipeline imputes missing values for property \textit{ET-GazeRight-Y} based on the the other three \textit{Gaze}-properties which are referenced directly by name. If one of these properties is renamed, the missing value imputation operator cannot access them and therefore the imputation does not work. Based on the implementation, the pipeline could crash. Differing data characteristics can lead to unexpected operator behavior as well. Thus, there is the need for a monitoring system which continuously inspects and assesses the data which is processed by the operators. This can help data engineers in tracking errors and also in finding root causes of errors more easily.

\begin{figure}[ht]
    \centering
\includegraphics[width=0.85\textwidth]{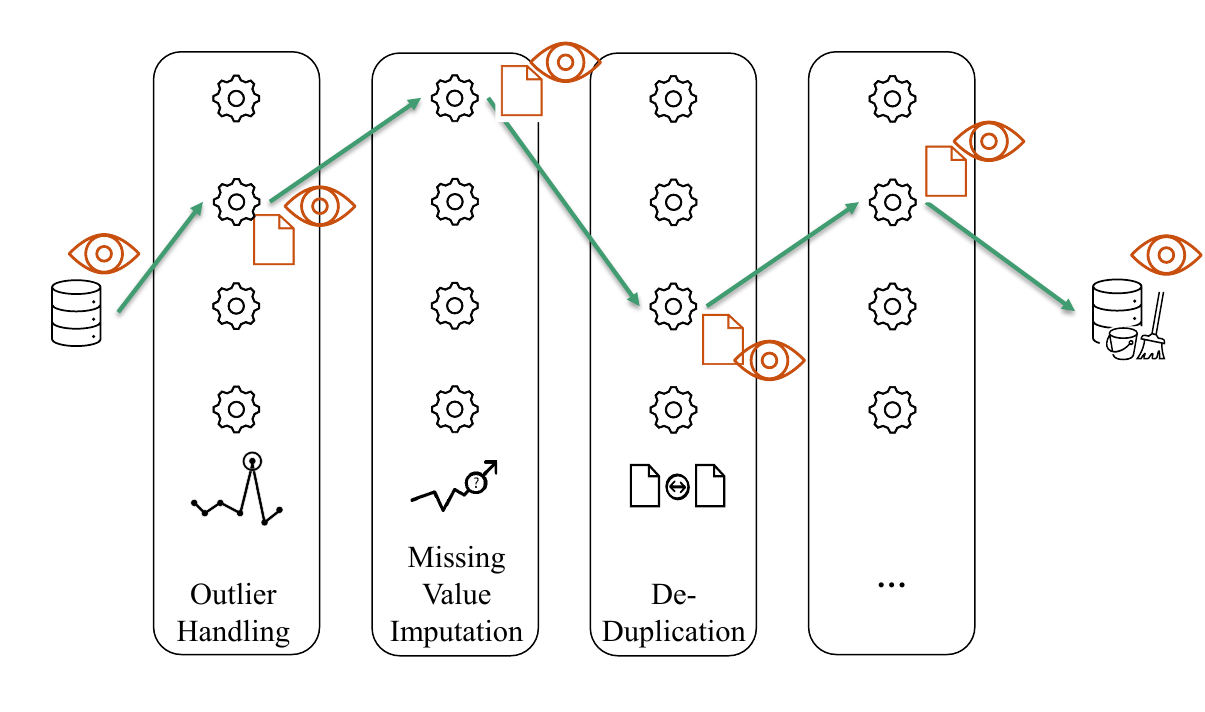}
	\caption{Integration of a data monitoring component into the pipeline. The orange symbols indicate the observation (= eye symbol) of operators by creating data profiles (= file symbol) of input, intermediate, and output data.}
    \label{fig:inspection}
\end{figure}

Figure~\ref{fig:inspection} shows how the monitoring component is integrated into the data pipeline. The pipeline takes as input the raw data and outputs the processed / clean data. It does this by instrumenting operators. These operators produce so-called \textit{intermediates}. Ideally, metadata about all data processed and produced in the workflow are tracked: input, intermediates, and output. Obviously, simply comparing raw input and output datasets in order to find unexpected changes is not sufficient due to the dimension and size of the data that are processed in modern data pipelines. Thus, we use the data profile we introduced in Section~\ref{sec:automatic_composition} which contains metadata about the processed data. We then determine significant changes by \textit{comparing the data profile} of the input with the data profile of the output. This provides \textit{end-to-end transparency} over the whole pipeline and helps in preventing failures and detecting unexpected changes to data and operators. Additionally, we plan to construct so-called \textit{data assertions}~\cite{gvdb_sebastian} which can be seen as denial constraints for new input data. They serve as a first means to detect data which is not suitable for processing by the pipeline. Common pipeline failures can be prevented this way. The data assertions are continuously updated based on new incoming data in order to adjust to changing data characteristics as well as possible and to prevent too many false warnings.

\subsubsection{Components of Data Profiles}
The data profile contains descriptive statistics about the data. As our goal is to cover a wide variety of different data models and use cases, we aim for a flexible way to define the fields of a data profile. The following paragraph exemplifies  possible metadata to be collected during pipeline runs. A data profile can be configured by the user by including a number of the subsequently displayed data characteristics.

We assume the input data to follow a certain structure. Therefore, an important first component of the data profile is the schema of the data set. Schema information includes property names and data types. As described before, we mainly differentiate between three data types the properties correspond to: numeric, categorical, and text. For numeric properties, descriptive statistics like  central tendency, variability, and shape can be derived. The amount and location of outliers are considered, too. In the case of categorical variables, statistics like frequencies, proportions, and mode are calculated from the data. Multiple statistics can also be generated for text data. Here, we focus on statistics which are not intended to capture semantic, but rather structural properties of the text. This includes the overall word count, word frequencies, or the vocabulary size. An additional approach -- which can be especially important in data analysis applications -- is profiling across multiple properties. Relationships and dependencies between properties can be found this way.  

\subsubsection{Generation of Data Profiles and Diffs}
Data profiles can be seen as a summary of a dataset. They not only serve as synopses of intermediates, but also as means to make intermediates processed in the data pipeline comparable. Changes in data conducted by operators can be made explicit by comparing two data profiles, therefore finding differences in data characteristics. The result of such a comparison is a \textit{data profile diff} which aims at providing useful information for users to understand and assess a certain data change conducted by operators or introduced in a new incoming batch. This way, changes deemed to be significant can be marked for data engineers. The idea of a data profile diff for comparing datasets was also introduced in our previous work~\cite{hilda}.

Our approach thus utilizes the two aforementioned main abstractions: data profiles and their diffs. Our envisioned self-aware data engineering pipeline system continuously calculates them during the deployment phase. The resulting data profiles and their diffs are then stored in order to enable a comparison between different pipeline runs. Differences in input datasets which are processed in two subsequent pipeline runs can be detected. By storing metadata over the whole deployment cycle of a data pipeline, it is also possible to track data drifts which are happening over a longer period of time. 
The same idea applies for data errors. Tracking the number of occurrences of various error types can also give important clues on changing data. An example would if a property which yielded no missing values at design time suddenly does deliver any values. This could indicate a broken data source. In order to not let such anomalies unnoticed, we also build an \textit{error profile diff}. 

In our running example, a data shift is shown in Figure~\ref{fig:et-dataset-semantic-change}. The distribution of the two properties \textit{Fixation-X} and \textit{Fixation-Y} has changed due to a screen change. However, a data drift could also have different root causes, e.g.\ the usage of different configurations on the eye tracker, the usage of another eye tracker, or the usage of a different measure for the fixation. Detecting such significant deviations and alerting the data engineer as early as possible can help in preventing resulting errors in downstream operators (e.g., crashes due to an unexpected input schema). It can also prevent those subsequent operators from outputting undesired results. In this example of a distribution shift for aforementioned properties, assume the pipeline optimizer (presented in Section~\ref{sec:automatic_composition}) initially chose to detect and consequently eliminate outliers by applying the Local Outlier Factor~\cite{Breunig2000}. However, now that the distribution changed, it is possible that better results could be achieved by applying a different outlier detection algorithm. Thus, there should be an evaluation if an \textit{adaptation} of the current pipeline is necessary. Ideas on how to conduct this evaluation and the subsequent adaptation are discussed in Section~\ref{sec:evolution}. 

\subsubsection{Research Challenges}
We identified multiple research challenges to be solved in order to achieve self-aware data engineering pipelines. The design of data profiles and their diffs is an important one in this context. Providing insights on how the data looks like and is changing throughout the pipeline or between different pipeline runs is the goal of these abstractions. Therefore, we aim for a high \textit{information value} i.e.\ a representation of metadata which helps data engineers the most when searching for root causes of errors or changes in the pipeline. For this, we expect an user survey to be necessary. Another interesting research question is how to incorporate user feedback into the data profile and diff generation to further improve it. 

A challenge which aims more for practicability is to minimize the computational overhead the monitoring system adds to the data engineering process. The metadata collection and the calculation of data profiles and data profile diffs should not influence the performance of the data engineering pipeline itself. Providing tools for efficiently querying information on pipeline runs is another important requirement to make the monitoring system feasible.

\bigskip

After achieving a self-aware data engineering pipeline, the system can now use this foundation to automatically react to change. \texttt{\textit{Level 3}} of our envisioned data engineering pyramid, depicted in Figure~\ref{fig:pyramid_v2}, contains self-adapting pipelines which automatically adapt to changes detected by the inspection system. In the next section, we take a closer look at our vision regarding this adaptation process.

\subsection{Level 3 : Self-adapting Data Engineering Pipeline}
\label{sec:evolution}

\begin{figure}[ht]
    \centering
    \includegraphics[width=0.90\textwidth]{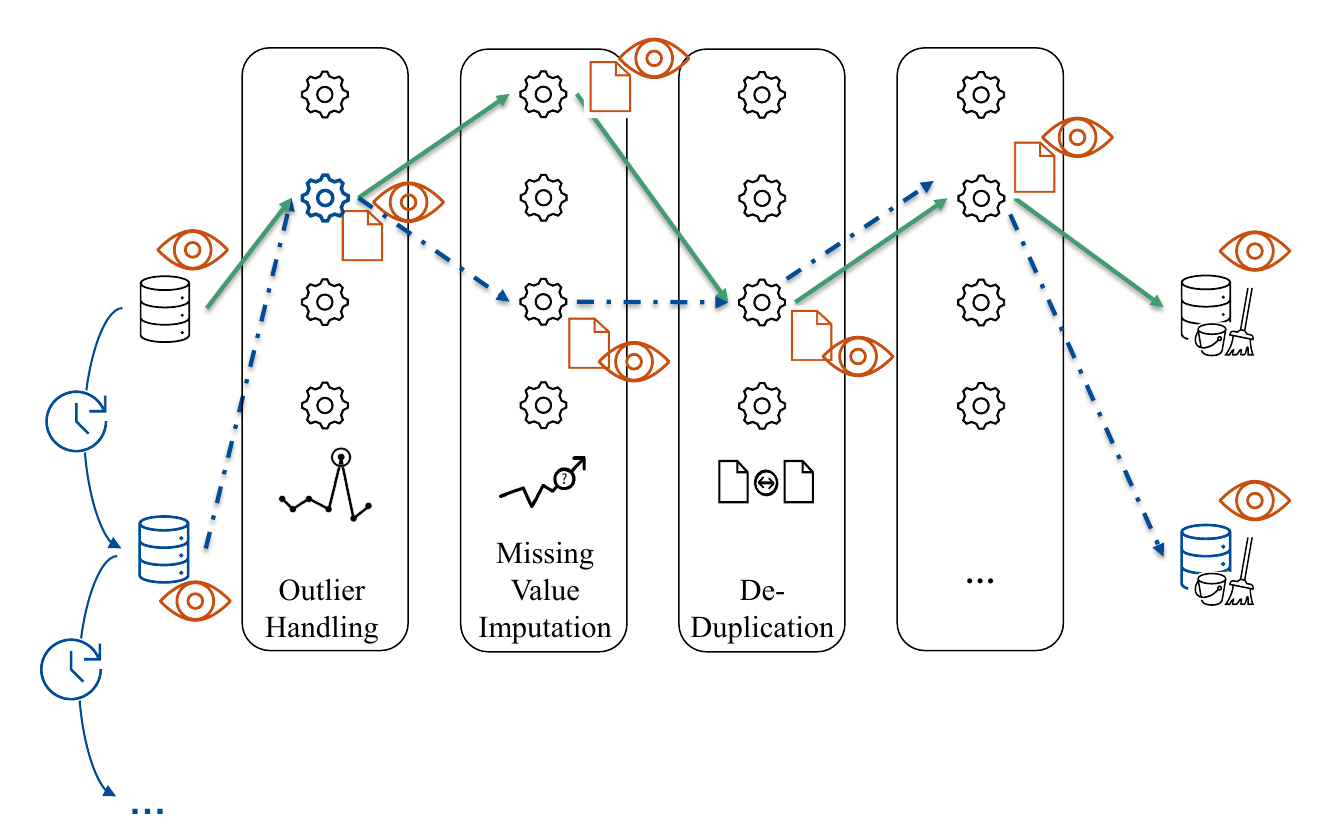}
	\caption{Self-adapting data engineering pipeline which automatically reacts to change over time (blue input data) by adapting its structure and components (dashed blue line and blue gear).}
    \label{fig:adaptation}
\end{figure}
Even though many of the parts needed for achieving self-adaptivity in software systems exist in part, as presented in Section~\ref{subsec:sota-level-3}, there is a need for a self-adapting data engineering pipeline, combining all these parts while assuring functionality and correctness. In \cite{DBLP:conf/gvd/Kramer23} we proposed conceptual requirements for such a system. The proposed solution from Sections~\ref{sec:automatic_composition} and~\ref{sec:inspection}, i.e.\ an optimal data engineering pipeline, which is monitored and inspected constantly, might still break. For instance, this could be due to a structural (see Figure~\ref{fig:et-dataset-structural-change}) or semantic change (see Figure~\ref{fig:et-dataset-semantic-change}) in the data. The inspection system, as presented in Section~\ref{sec:inspection} will notice these types of evolution and could alert a data engineer in order to fix these issues. Involving a human in this step is expensive and drives up maintenance cost during the lifecycle of a data engineering pipeline. At the same time it is not clear, if, when, and to what extent such a change might occur. Therefore, trying to achieve a high level of automation appears to be very beneficial. This shows the need for a system which not only generates an optimal pipeline initially and detects significant change through constant monitoring and inspection, but which automatically adapts to said changes. 

Figure~\ref{fig:adaptation} showcases the vision of such a self-adapting pipeline. Handling evolutionary change can take different forms, e.g.\ adjusting the configuration of an operator to work with the new circumstances or changing the structure of the pipeline by adding or removing a component. Even though many diverse forms of change and corresponding adaptation exist, this section focuses on changes regarding semantics and schema of data, their impact on the data engineering pipeline, and possible adaptation approaches.

As mentioned before, the schema of data is very important and the first part of a data profile. Schemas can be considered as a layer between the data and the pipeline, used by operators to correctly access values. If a schema changes over time without adapting the operators utilizing it, errors will occur. Semantic changes will also produce errors if they are not adapted. We envision diverse types of \textit{failures} associated with the different categories of change, each needing their own form of adaptations.

For structural and semantic change, two types of failures arise: (1) At least one operator of the pipeline is not functional anymore leading to a runtime error. (2) The pipeline is still functional but is not semantically correct anymore, i.e.\ at least one operator produces semantically incorrect results. If an operator is confronted with a changed schema and is still functional we call this operator \textit{schema robust}. There are multiple levels of schema robustness. Each new level builds upon the previous one. The base level is indifference towards certain change. For example, if the schema with respect to a specific property is changed, all operators not working on said property will not run into any problems or errors. The next schema robustness level is associated with loss of information, i.e.\ the schema of the current batch is a proper subset of the preceding batch's schema. The next level deals with information change, occurring for example through renaming of a property. The final level of schema robustness comes into play, when new information is added, i.e.\ the current schema is a proper superset of the preceding batch's schema. This is the case, when a new property is added between batches. Besides indifference to change, we need self-adaptation capabilities to some degree, in order to achieve robustness for the three remaining levels of schema robustness. The schema changes need to be analyzed and operators need to be adjusted based on the inferred results. In~\cite{robustness}, we introduce a decision tree to assess whether a specific operator remains robust against changes caused by a particular SMO or requires self-adaptation.

Semantic change always includes a context founded in the data itself, in which the assessment of change happens. For example, missing value imputation by imputing the mean value might use the mean value of the current batch or calculate a mean value based on the current batch and historic batches. Only in the latter case a semantic shift is possible. Therefore we conclude, that \textit{semantic robustness} is significantly dependent on the semantic context of an operator. Generally speaking, robustness is never assured and adapting the pipeline to changing circumstances might become necessary.

We combine schema and semantic robustness to define \textit{operator robustness}. However, it is clear to us, that the schema and semantic dimensions are not sufficient for a complete definition of operator robustness. 

When a change in the schema or in the semantics of data is found by the self-aware pipeline as proposed in Section~\ref{sec:inspection} the self-adapting pipeline takes over. Presented with the profile diff containing all significant changes, the adaptation process starts. It can be broadly structured into three phases, resembling parts of the MAPE-K (monitor, analyze, plan, execute, knowledge) architecture as proposed in \cite{MAPE-K}:

\begin{enumerate}
    \item \textbf{Change Interpretation}: Interpreting the changes and breaking them down into independent steps
    \item \textbf{Adaptation Analysis}: Selecting fitting adaptations as well as identifying the affected pipeline parts
    \item \textbf{Propagation and Evaluation}: Propagating the adaptation to the actual pipeline and evaluating it
\end{enumerate}

The three phases are detailed in the following.
\subsubsection{Change Interpretation}
The first phase in adapting a data engineering pipeline in the context of significant change is to analyse the changes. It is possible for consecutive data batches to include more than one change, e.g.\ a changed schema and a shift in a properties value distribution. It is important to analyze these changes in order to break them down and to form manageable chunks for adaptation. During this process it is important to resolve ambiguities, for example in the schema layer~\cite{DBLP:conf/bigdataconf/KlettkeAS0S17} and to find dependencies between individual changes. The former can be achieved by using the available metadata to contextualize ambiguities and therefore resolve them. An example for the latter is a semantic shift which might have occurred to a property whose name was also changed, i.e.\ a rename operation was applied to the schema. Take for instance our running example and assume that both semantic (Figure~\ref{fig:et-dataset-semantic-change}) and structural (Figure~\ref{fig:et-dataset-structural-change}) change to property \textit{Fixation-X} happened between consecutive batches. In this case, both changes will affect a similar if not the same set of operators. At the same time, it is important to first adapt the rename from \textit{Fixation-X} to \textit{Fixation-Screen-X} followed by addressing the semantic change, otherwise the semantic context will not be correct for operators utilizing it. The goal of this first phase is to find all such dependencies and form a set of independent \textit{change steps} to which a parallel set of \textit{adaptation steps} can be constructed.

\subsubsection{Adaptation Analysis}
Given all independent change steps, the second adaptation phase starts with creating a search space of all possible adaptation operations. In some cases, there is only one such operation, e.g.\ a renamed attribute is adapted by applying the new name to operators working on it, for instance after renaming \textit{Fixation-X} to \textit{Fixation-Screen-X} as shown in Figure~\ref{fig:et-dataset-structural-change} of the running example. In other cases, more options exist, e.g.\ changing an operator altogether or only adjusting its configuration after a property's distribution changed. 

Diverse possibilities for adaptation arise, when a new property is added, because it leads to completely new information, i.e.\ data and metadata, which needs to be contextualized first, in order to adapt the pipeline to it and ensure not only functionality but also semantic correctness with respect to data quality. This contextualization can be achieved in different ways. Some approaches and corresponding examples are shown in the following list. 

\begin{itemize}
    \item Heuristics: Utilizing metadata
    \item Statistics: Calculating correlations
    \item Machine learning: Training and using a classifier
    \item Knowledge graphs: Representing and finding semantic connections
    \item Large language models: Prompt engineering for solving ambiguities
\end{itemize}

Simple heuristics based on readily available metadata can be applied, e.g.\ adaptation based on the data type. More complex techniques also appear promising, such as utilizing knowledge graphs of the given domain. Their nodes and edges can be linked to the data and their relationships. This allows for inference of new information, i.e.\ properties, leveraging the semantic connections provided by the knowledge graph \cite{Zou2020ASO}. Capitalizing on the reasoning capabilities of large language models these systems might also be used to solve the given task as some preliminary results suggest \cite{mior2024largelanguagemodelsjson, fu2024compoundschemaregistry}. 

Even after contextualizing all available information, it is still possible that a lot of potential adaptations remain. Therefore it is advisable to use some form of best practices, which provide a framework for groups of highly similar adaptations \cite{DBLP:conf/summersoc/GlakeKSR21}. In this way a set containing all adaptation steps is created. In conjunction with this process, we also need to find the locations, i.e.\ algorithms and their parameters which are affected by the change and therefore need to be adapted.
To solve this, we envision a \textit{profile registry} including a \textit{schema version graph}. Both these components will be explained in detail in Section~\ref{sec:system}. All needed information about operators and their parameters, the schema, the data and the pipeline itself is stored there. Especially the pipeline profile as presented in Scetion~\ref{sec:auto_comp:reschal} is an important interface for adaptation. It holds the definition of the current pipeline, which can be altered with respect to a specific adaptation step.

For instance, assume that an operator works on property \textit{Fixation-X} as shown in Figure~\ref{fig:et-dataset-structural-change} and is now confronted with a rename. The pipeline profile can then be altered to use the new name instead of \textit{propertyName = Fixation-X}. The change between the old pipeline profile and the new one, which includes the renamed property, can again be expressed in the form of a profile diff.

In some cases, change might be so dramatic, that a self-adaptation by the system is not feasible anymore. In these situations, it is better to rerun the automatic pipeline composition as presented in Section~\ref{sec:automatic_composition}. 

\subsubsection{Propagation and Evaluation}
The final phase of adaptation encompasses the \textit{propagation}\/ of the adaptations to the actual pipeline and its evaluation. Depending on the architecture and technologies used for the pipeline this can take different forms. An important aspect to consider here is whether or not the pipeline is constantly running as a service, waiting for new data to arrive, or if it only runs when triggered, for example by an orchestrator. Changes to the code, e.g.\ by means of automatic code generation, can only be applied to pipelines not running as a service. In either case, changing the code can also lead to inefficiencies, when the programming language that is used for the pipeline needs to be compiled. In contrast, every form of change and corresponding adaptation can be resolved through changing the code. Using a mapping layer as briefly described above, eases adaptation. Schema definition files or operator contracts can always be changed, since they are saved separately from the code and can be dynamically loaded, whenever the associated operator fires, even when the pipeline is run as a service.

For these different cases, we envision specific types of adapters as mentioned in Section~\ref{sec:auto_comp:reschal}, which bridge the gap between our abstract self-adaptive system and the concrete data engineering pipeline. For example, the rename operation mentioned above, which leads to a change in the pipeline profile (diff), could be resolved through the adjustment of the property name in a schema mapping file, through sending API calls to a workflow system or by utilizing dynamic variable name resolution through meta-programming.

Once all adaptation steps have been applied the pipeline can be run and needs to be \textit{evaluated}. Evaluating an adaptation can be done in the context of a given optimization metric. In our case data quality is used, but other primary or secondary optimization goals appear valid, such as performance~\cite{CheDDaR}. Evaluating the functionality of an adapted pipeline is trivial, evaluating the semantic correctness is very hard in comparison. Functionality is ensured if the pipeline runs without breaking. Semantic correctness is, again, based on a specific context. In a fully automatic self-adapting pipeline the contextualization as described above produces more or less probable results, i.e.\ adaptations and resulting pipelines. The evaluation of semantic correctness correlates strongly with this probability.

\subsubsection{Research Challenges}
Achieving the vision of self-adapting data engineering pipelines encompasses a plethora of research challenges. A key challenge is contextualizing existing or new information respectively in order to find a suitable adaptation to the new circumstances. Finding the right methods or ensemble of methods for specific types of information is not trivial. Deriving best practices for groups of similar adaptations also falls in this area and is also an important goal for making the adaptation process feasible. For evaluation purposes and to gain empirically sound insight, a benchmarking environment for testing contextualizations and their associated adaptations is needed. This is especially important to evaluate semantic correctness after an adaptation.

Another challenge will be the modeling of an abstract data engineering pipeline representation. Such a structure is needed in order to combine and store all types of data, metadata and information on the whole data engineering pipeline and the process. This includes developing a suitable interface, i.e.\ an API, which not only allows system components to exchange information seamlessly, but is also accessible to humans. A graph database might be a promising candidate for this component, but other tools and technologies have to be considered. This also leads to general inquiries into software architecture, programming paradigms, and technologies, which have to be made in order to find the ones which support the adaptation process the most.

A significant challenge will be to widen the scope, not only to other adaptations concerning schema and semantic changes, but also to other areas such as changing operators or runtime environments.

\section{Proposed System}~\label{sec:system}
\begin{figure}[ht]
    \centering
    \includegraphics[width=\textwidth]{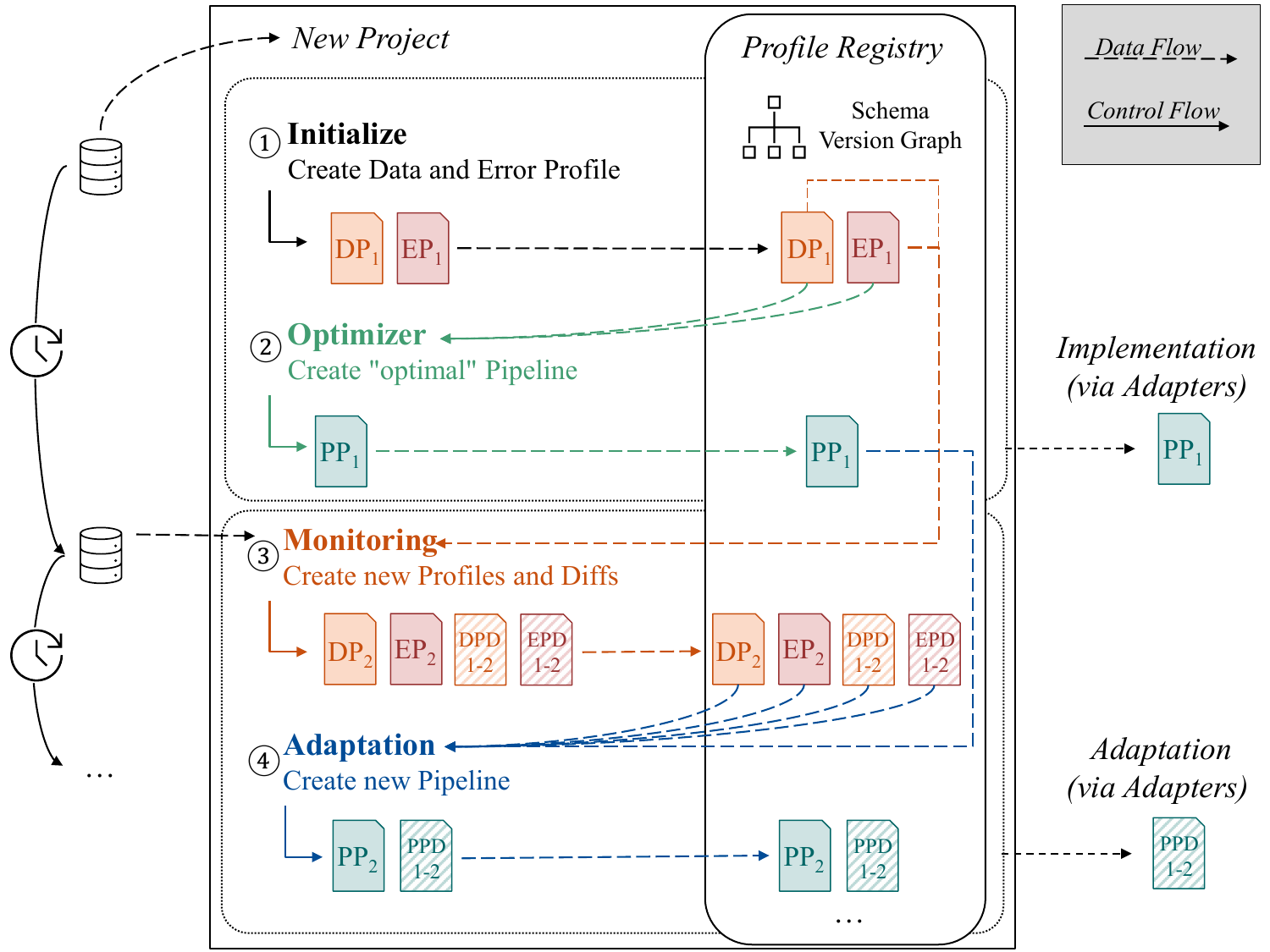}
	\caption{Overview of the proposed system and its components \newline (\textbf{DP} = data profile, \textbf{EP} = error profile, \textbf{PP} = pipeline profile, \textbf{DPD} = data profile diff, \textbf{EPD} = error profile diff, \textbf{PPD} = pipeline profile diff)}
    \label{fig:proposed-system}
\end{figure}

Our proposed system is described in the following. It comprises the three levels presented. To this end, different profiles are a central aspect as described. By using profiles and diffs between profiles, the actual data records are only ever required initially. As described in Section~\ref{sec:automatic_composition}, the system is abstracted from the technology of the specific pipeline. The implementation is carried out via adapters. This makes the system generic and reusable. Figure~\ref{fig:proposed-system} shows  our proposed system and the use of the different profiles in this context.

For a new use case, the initial data batch is loaded into the system. A new project is created there first. Each project has an entry in the \textit{profile registry}. This is a core component of our proposed system, which stores all metadata, i.e.\, all profiles, which are created throughout the process. The necessity of a storage component was already discussed in Section~\ref{sec:inspection}. It is used to monitor pipeline performance on different batches of data. This way, data changes over time can be detected. Furthermore, it helps on evaluating how these changes affect the pipeline. The profile registry in conjunction with a \textit{schema version graph}\/ also allows for data and workflow provenance, i.e.\ traceability of data, pipeline operations and results.

\paragraph{\tikzcircle{1} Initialization}
The first step is an \textit{initialization}, where the \textit{data profile} (DP) and the \textit{error profile} (EP) are created from the input data.

As described in Sections~\ref{sec:automatic_composition} and~\ref{sec:inspection}, a data profile is a concrete set of metadata to describe a data set. The error profile, on the other hand, contains all the information about the errors in the data set. This is needed to identify the errors to be cleaned and to decide which cleaning operators are required. It is created with CheDDaR~\cite{CheDDaR, CheDDaR2025}, as explained in Section~\ref{sec:automatic_composition}. 

In addition, the initial schema is saved in the \textit{schema version graph} using a schema extraction algorithm~\cite{DBLP:conf/btw/KlettkeSS15,DBLP:conf/bigdataconf/KlettkeAS0S17}. This initial schema is required later in the monitoring and adaptation components to determine the changes.

\paragraph{\tikzcircle{2} Optimizer}
Data and error profile are then used by the \textit{optimizer} to generate the best possible pipeline. It is transferred to an adapter in the form of a \textit{pipeline profile} (PP)~\cite{ALPINE}. This represents an abstract description of the pipeline as described in Sections~\ref{sec:automatic_composition} and~\ref{sec:evolution}. Using adapters, a concrete pipeline can then be generated from this profile in a technology of choice. One adapter we plan to implement generates a Python script which operates on pandas DataFrames and executes data processing steps based on popular data science libraries like \textit{pandas} and \textit{scikit-learn}. Other adapters could for instance generate a DAG for Apache Airflow which implements the defined processing steps. 

\paragraph{\tikzcircle{3} Monitoring}
If a new batch of data arrives after some time, it is passed to the \textit{monitoring} component, which generates the new data and error profiles. In addition to the data profile itself, a \textit{data profile diff} (DPD) is created, as described in Section~\ref{sec:inspection}. 

Besides storing the changes in the actual data values, e.g.\ a distribution shift for a given property, the data profile diff also stores all new and deleted schema elements. Building upon our prior work~\cite{DBLP:conf/bigdataconf/KlettkeAS0S17}, schema element differences are used as input for a \textit{schema modification operation} (SMO) inference algorithm. Utilizing the extracted SMOs, the schema version graph, which at this point only stores the initial schema, is extended. This graph is an important component for self-adaptation with respect to structural changes. For instance, it can be used to determine whether or not a new schema has been used in a prior batch, allowing the system to reuse the associated pipeline without self-adaptation.

Accordingly, in addition to the error profile, an \textit{error profile diff} (EPD) is created. This can be necessary if a new batch of data contains different errors than the batches before.

\paragraph{\tikzcircle{4} Adaptation}
If a significant change is detected, the \textit{adaptation} is performed. This uses the data and error profile diffs as well as the pipeline profile and creates a new pipeline profile as well as a \textit{pipeline profile diff} (PPD). There, the changes to the existing pipeline are documented as described in Section~\ref{sec:evolution}. If an adaptation is necessary, adapters can again be used to make changes to the specific pipeline based on this pipeline profile diff.

\paragraph{Processing of further batches}
The process is the same for further batches: Skipping steps \tikzcircle{1} and \tikzcircle{2}, the data is transferred to the \tikzcircle{3} \textit{monitoring} component, which creates data and error profiles as well as the associated diffs and checks for changes. If necessary, an \tikzcircle{4} \textit{adaptation} is made. The \tikzcircle{2} \textit{optimizer} is only executed again if there are such serious changes in the data that an adaptation is not usefully applicable. This then creates a new pipeline in the form of a pipeline profile.

\bigskip

Generally speaking, we aim to create a system, which is as autonomous as possible. Maintenance is a significant cost factor for long-running software projects. Reducing it through automation could be very beneficial in this regard. At the same time, full automation is not desirable or feasible in certain situations. Therefore, we intend to include interfaces for human interaction during all stages of our proposed system. Especially, when choices made by the system are probabilistic, a final decision made by a human can become necessary. This is the case in circumstances, where responsibility and accountability with respect to the system result is crucial. Example areas for this include finance and healthcare.

\section{Conclusion}~\label{sec:conclusion}
In this paper, we presented our vision of next generation data engineering pipelines. These consist of several levels. \texttt{\textit{Level 1}} represents an optimized data engineering pipeline. Optimization in this case refers to data quality, which means that a pipeline is designed that leads to the best possible data quality. Once this pipeline is deployed in production, \texttt{\textit{Level 2}} -- the self-aware data engineering pipeline -- commences. This automatically recognizes changes in the data and intermediate states of the pipeline. It thus enables a more targeted root cause analysis of errors and creates more transparency throughout the entire process. The self-aware pipeline also provides the basis for \texttt{\textit{Level 3}} -- the self-adapting pipeline. This can react automatically to detected changes in the data and adapt the pipeline. In this way, aborts or incorrect results of the pipeline are prevented. If all three levels are fulfilled, an autonomous system is achieved that constantly monitors the current status, reacts to changes, and hence leads to the best possible data quality at all times.

We proposed such a system and identified the required components. It can be used to create, monitor and adapt the optimal pipeline. By using pipeline profiles and their diffs, the system can be used flexibly, regardless of the technology the pipeline is implemented with.

Furthermore, we have described the associated research challenges for each of the three levels. These must be addressed in future work. One of these is particularly important, as it is required throughout all components of our envisioned autonomous data engineering pipeline: an extensive data profiling. For pipeline optimization, data profiles are used for rule-based optimization to identify suitable algorithms. In the context of monitoring and potentially adapting the pipeline, profiling the data, and especially detecting changes between batches of data is crucial in order to provide useful monitoring tools and enable sound contextualization and thereby effective adaptation. Consequently, we will develop and evaluate different approaches for profiling data and determining deviations between datasets which can be used in the setting of data engineering pipelines.

\section*{Acknowledgment}
Many thanks to Jennifer Landes for providing the data used in the running example.

\bibliographystyle{unsrtnat}
\bibliography{references}

\begin{thebibliography}{83}
\providecommand{\natexlab}[1]{#1}
\providecommand{\url}[1]{\texttt{#1}}
\expandafter\ifx\csname urlstyle\endcsname\relax
  \providecommand{\doi}[1]{doi: #1}\else
  \providecommand{\doi}{doi: \begingroup \urlstyle{rm}\Url}\fi

\bibitem[Klettke and St{\"{o}}rl(2022)]{4Generations}
Meike Klettke and Uta St{\"{o}}rl.
\newblock {Four Generations in Data Engineering for Data Science}.
\newblock \emph{Datenbank-Spektrum}, 22\penalty0 (1):\penalty0 59--66, 2022.
\newblock \doi{10.1007/S13222-021-00399-3}.

\bibitem[Cai and Zhu(2015)]{Cai2015}
Li~Cai and Yangyong Zhu.
\newblock {The Challenges of Data Quality and Data Quality Assessment in the Big Data Era}.
\newblock \emph{Data Sci. J.}, 14:\penalty0 2, 2015.
\newblock \doi{10.5334/DSJ-2015-002}.

\bibitem[Restat et~al.(2022)Restat, Boerner, Conrad, and St{\"{o}}rl]{GouDa}
Valerie Restat, Gerrit Boerner, Andr{\'{e}} Conrad, and Uta St{\"{o}}rl.
\newblock {GouDa - generation of universal data sets: improving analysis and evaluation of data preparation pipelines}.
\newblock In \emph{DEEM@SIGMOD}, pages 2:1--2:6. {ACM}, 2022.
\newblock \doi{10.1145/3533028.3533311}.

\bibitem[Ilyas and Chu(2019)]{Ilyas2019}
Ihab~F. Ilyas and Xu~Chu.
\newblock \emph{{Data Cleaning}}, volume~28 of \emph{{ACM} Books}.
\newblock {ACM}, 2019.
\newblock \doi{10.1145/3310205}.

\bibitem[Tebernum et~al.(2021)Tebernum, Altendeitering, and Howar]{DERM}
Daniel Tebernum, Marcel Altendeitering, and Falk Howar.
\newblock {DERM:} {A} reference model for data engineering.
\newblock In \emph{{DATA}}, pages 165--175. {SCITEPRESS}, 2021.
\newblock \doi{10.5220/0010517301650175}.

\bibitem[Matskin et~al.(2021)Matskin, Tahmasebi, Layegh, Payberah, Thomas, Nikolov, and Roman]{Matskin2021ASO}
Mihhail Matskin, Shirin Tahmasebi, Amirhossein Layegh, Amir~Hossein Payberah, Aleena Thomas, Nikolay Nikolov, and Dumitru Roman.
\newblock {A Survey of Big Data Pipeline Orchestration Tools from the Perspective of the DataCloud Project}.
\newblock In \emph{{DAMDID/RCDL} (Supplementary Proceedings)}, volume 3036 of \emph{{CEUR} Workshop Proceedings}, pages 63--78. CEUR-WS.org, 2021.
\newblock URL \url{https://ceur-ws.org/Vol-3036/paper05.pdf}.

\bibitem[Petrovska(2021)]{Petrovska2021}
Ana Petrovska.
\newblock {Self-Awareness as a Prerequisite for Self-Adaptivity in Computing Systems}.
\newblock In \emph{{ACSOS-C}}, pages 146--149. {IEEE}, 2021.
\newblock \doi{10.1109/ACSOS-C52956.2021.00039}.

\bibitem[Landes et~al.(2024)Landes, K{\"{o}}ppl, and Klettke]{et_jenny}
Jennifer Landes, Sonja K{\"{o}}ppl, and Meike Klettke.
\newblock {Data Processing Pipeline for Eye-Tracking Analysis}.
\newblock In \emph{GvDB}, volume 3710 of \emph{{CEUR} Workshop Proceedings}, pages 35--42. CEUR-WS.org, 2024.
\newblock URL \url{https://ceur-ws.org/Vol-3710/paper6.pdf}.

\bibitem[Koupil and Holubov{\'{a}}(2022)]{Koupil2022}
Pavel Koupil and Irena Holubov{\'{a}}.
\newblock {A unified representation and transformation of multi-model data using category theory}.
\newblock \emph{J. Big Data}, 9\penalty0 (1):\penalty0 61, 2022.
\newblock \doi{10.1186/S40537-022-00613-3}.

\bibitem[Romero and Wrembel(2020)]{DBLP:conf/dawak/RomeroW20}
Oscar Romero and Robert Wrembel.
\newblock {Data Engineering for Data Science: Two Sides of the Same Coin}.
\newblock In \emph{DaWaK}, volume 12393 of \emph{Lecture Notes in Computer Science}, pages 157--166. Springer, 2020.
\newblock \doi{10.1007/978-3-030-59065-9\_13}.

\bibitem[Biswas et~al.(2022)Biswas, Wardat, and Rajan]{DBLP:conf/icse/BiswasWR22}
Sumon Biswas, Mohammad Wardat, and Hridesh Rajan.
\newblock {The Art and Practice of Data Science Pipelines: {A} Comprehensive Study of Data Science Pipelines In Theory, In-The-Small, and In-The-Large}.
\newblock In \emph{{ICSE}}, pages 2091--2103. {ACM}, 2022.
\newblock \doi{10.1145/3510003.3510057}.

\bibitem[Heck(2024)]{AI-DE}
Petra Heck.
\newblock {What About the Data? {A} Mapping Study on Data Engineering for {AI} Systems}.
\newblock In \emph{{CAIN}}, pages 43--52. {ACM}, 2024.
\newblock \doi{10.1145/3644815.3644954}.

\bibitem[Chang and Grady(2019)]{BigData-4Vs}
Wo~Chang and Nancy Grady.
\newblock {NIST Big Data Interoperability Framework: Volume 1, Definitions}.
\newblock Technical report, National Institute of Standards and Technology, 2019.

\bibitem[Volk et~al.(2019)Volk, Staegemann, Pohl, and Turowski]{BigData-DE}
Matthias Volk, Daniel Staegemann, Matthias Pohl, and Klaus Turowski.
\newblock {Challenging Big Data Engineering: Positioning of Current and Future Development}.
\newblock In \emph{IoTBDS}, pages 351--358. SciTePress, 2019.
\newblock \doi{10.5220/0007748803510358}.

\bibitem[Liew et~al.(2017)Liew, Atkinson, Galea, Ang, Martin, and van Hemert]{DBLP:journals/csur/LiewAGA0H17}
Chee~Sun Liew, Malcolm~P. Atkinson, Michelle Galea, Tan~Fong Ang, Paul Martin, and Jano~I. van Hemert.
\newblock {Scientific Workflows: Moving Across Paradigms}.
\newblock \emph{{ACM} Comput. Surv.}, 49\penalty0 (4):\penalty0 66:1--66:39, 2017.
\newblock \doi{10.1145/3012429}.

\bibitem[Santu et~al.(2022)Santu, Hassan, Smith, Xu, Zhai, and Veeramachaneni]{Karmaker2022}
Shubhra Kanti~Karmaker Santu, Md.~Mahadi Hassan, Micah~J. Smith, Lei Xu, Chengxiang Zhai, and Kalyan Veeramachaneni.
\newblock {{AutoML} to Date and Beyond: Challenges and Opportunities}.
\newblock \emph{{ACM} Comput. Surv.}, 54\penalty0 (8):\penalty0 175:1--175:36, 2022.
\newblock \doi{10.1145/3470918}.

\bibitem[Xin et~al.(2021)Xin, Wu, Lee, Salehi, and Parameswaran]{Xin2021}
Doris Xin, Eva~Yiwei Wu, Doris Jung~Lin Lee, Niloufar Salehi, and Aditya~G. Parameswaran.
\newblock {Whither {AutoML}? Understanding the Role of Automation in Machine Learning Workflows}.
\newblock In \emph{{CHI}}, pages 83:1--83:16. {ACM}, 2021.
\newblock \doi{10.1145/3411764.3445306}.

\bibitem[Krishnan et~al.(2017)Krishnan, Franklin, Goldberg, and Wu]{boostclean}
Sanjay Krishnan, Michael~J. Franklin, Ken Goldberg, and Eugene Wu.
\newblock {BoostClean: Automated Error Detection and Repair for Machine Learning}.
\newblock \emph{CoRR}, abs/1711.01299, 2017.
\newblock URL \url{http://arxiv.org/abs/1711.01299}.

\bibitem[Rekatsinas et~al.(2017)Rekatsinas, Chu, Ilyas, and R{\'{e}}]{holoclean}
Theodoros Rekatsinas, Xu~Chu, Ihab~F. Ilyas, and Christopher R{\'{e}}.
\newblock {HoloClean: Holistic Data Repairs with Probabilistic Inference}.
\newblock \emph{Proc. {VLDB} Endow.}, 10\penalty0 (11):\penalty0 1190--1201, 2017.
\newblock \doi{10.14778/3137628.3137631}.

\bibitem[Mahdavi and Abedjan(2020)]{baran}
Mohammad Mahdavi and Ziawasch Abedjan.
\newblock {Baran: Effective Error Correction via a Unified Context Representation and Transfer Learning}.
\newblock \emph{Proc. {VLDB} Endow.}, 13\penalty0 (11):\penalty0 1948--1961, 2020.
\newblock URL \url{http://www.vldb.org/pvldb/vol13/p1948-mahdavi.pdf}.

\bibitem[Chu et~al.(2015)Chu, Morcos, Ilyas, Ouzzani, Papotti, Tang, and Ye]{katara}
Xu~Chu, John Morcos, Ihab~F. Ilyas, Mourad Ouzzani, Paolo Papotti, Nan Tang, and Yin Ye.
\newblock {{KATARA:} {A} Data Cleaning System Powered by Knowledge Bases and Crowdsourcing}.
\newblock In \emph{{SIGMOD} Conference}, pages 1247--1261. {ACM}, 2015.
\newblock \doi{10.1145/2723372.2749431}.

\bibitem[Hameed and Naumann(2020)]{Hameed2020}
Mazhar Hameed and Felix Naumann.
\newblock {Data Preparation: {A} Survey of Commercial Tools}.
\newblock \emph{{SIGMOD} Rec.}, 49\penalty0 (3):\penalty0 18--29, 2020.
\newblock \doi{10.1145/3444831.3444835}.

\bibitem[Abedjan et~al.(2016)Abedjan, Chu, Deng, Fernandez, Ilyas, Ouzzani, Papotti, Stonebraker, and Tang]{Abedjan2016}
Ziawasch Abedjan, Xu~Chu, Dong Deng, Raul~Castro Fernandez, Ihab~F. Ilyas, Mourad Ouzzani, Paolo Papotti, Michael Stonebraker, and Nan Tang.
\newblock {Detecting Data Errors: Where are we and what needs to be done?}
\newblock \emph{Proc. {VLDB} Endow.}, 9\penalty0 (12):\penalty0 993--1004, 2016.
\newblock \doi{10.14778/2994509.2994518}.

\bibitem[Siddiqi et~al.(2023)Siddiqi, Kern, and Boehm]{Siddiqi2023}
Shafaq Siddiqi, Roman Kern, and Matthias Boehm.
\newblock {{SAGA:} {A} Scalable Framework for Optimizing Data Cleaning Pipelines for Machine Learning Applications}.
\newblock \emph{Proc. {ACM} Manag. Data}, 1\penalty0 (3):\penalty0 218:1--218:26, 2023.
\newblock \doi{10.1145/3617338}.

\bibitem[Boehm et~al.(2020)Boehm, Antonov, Baunsgaard, Dokter, Ginth{\"{o}}r, Innerebner, Klezin, Lindstaedt, Phani, Rath, Reinwald, Siddiqui, and Wrede]{Boehm2020}
Matthias Boehm, Iulian Antonov, Sebastian Baunsgaard, Mark Dokter, Robert Ginth{\"{o}}r, Kevin Innerebner, Florijan Klezin, Stefanie~N. Lindstaedt, Arnab Phani, Benjamin Rath, Berthold Reinwald, Shafaq Siddiqui, and Sebastian~Benjamin Wrede.
\newblock {{SystemDS}: {A} Declarative Machine Learning System for the End-to-End Data Science Lifecycle}.
\newblock In \emph{{CIDR}}. www.cidrdb.org, 2020.
\newblock URL \url{http://cidrdb.org/cidr2020/papers/p22-boehm-cidr20.pdf}.

\bibitem[Neutatz et~al.(2021)Neutatz, Chen, Abedjan, and Wu]{Neutatz2021}
Felix Neutatz, Binger Chen, Ziawasch Abedjan, and Eugene Wu.
\newblock {From Cleaning before {ML} to Cleaning for {ML}}.
\newblock \emph{{IEEE} Data Eng. Bull.}, 44\penalty0 (1):\penalty0 24--41, 2021.
\newblock URL \url{http://sites.computer.org/debull/A21mar/p24.pdf}.

\bibitem[Sambasivan et~al.(2021)Sambasivan, Kapania, Highfill, Akrong, Paritosh, and Aroyo]{Sambasivan2021}
Nithya Sambasivan, Shivani Kapania, Hannah Highfill, Diana Akrong, Praveen~K. Paritosh, and Lora Aroyo.
\newblock {Everyone wants to do the model work, not the data work": Data Cascades in High-Stakes {AI}}.
\newblock In \emph{{CHI}}, pages 39:1--39:15. {ACM}, 2021.
\newblock \doi{10.1145/3411764.3445518}.

\bibitem[Wang et~al.(1993)Wang, Kon, and Madnick]{Wang1993}
Richard~Y. Wang, Henry~B. Kon, and Stuart~E. Madnick.
\newblock {Data Quality Requirements Analysis and Modeling}.
\newblock In \emph{{ICDE}}, pages 670--677. {IEEE} Computer Society, 1993.
\newblock \doi{10.1109/ICDE.1993.344012}.

\bibitem[Wang et~al.(1995)Wang, Storey, and Firth]{Wang1995}
Richard~Y. Wang, Veda~C. Storey, and Christopher~P. Firth.
\newblock {A Framework for Analysis of Data Quality Research}.
\newblock \emph{{IEEE} Trans. Knowl. Data Eng.}, 7\penalty0 (4):\penalty0 623--640, 1995.
\newblock \doi{10.1109/69.404034}.

\bibitem[Blake and Mangiameli(2011)]{Blake2011}
Roger~H. Blake and Paul Mangiameli.
\newblock {The Effects and Interactions of Data Quality and Problem Complexity on Classification}.
\newblock \emph{{ACM} J. Data Inf. Qual.}, 2\penalty0 (2):\penalty0 8:1--8:28, 2011.
\newblock \doi{10.1145/1891879.1891881}.

\bibitem[Heinrich et~al.(2018)Heinrich, Hristova, Klier, Schiller, and Szubartowicz]{Heinrich2018}
Bernd Heinrich, Diana Hristova, Mathias Klier, Alexander Schiller, and Michael Szubartowicz.
\newblock {Requirements for Data Quality Metrics}.
\newblock \emph{{ACM} J. Data Inf. Qual.}, 9\penalty0 (2):\penalty0 12:1--12:32, 2018.
\newblock \doi{10.1145/3148238}.

\bibitem[Schelter et~al.(2018)Schelter, Lange, Schmidt, Celikel, Bie{\ss}mann, and Grafberger]{Schelter2018}
Sebastian Schelter, Dustin Lange, Philipp Schmidt, Meltem Celikel, Felix Bie{\ss}mann, and Andreas Grafberger.
\newblock {Automating Large-Scale Data Quality Verification}.
\newblock \emph{Proc. {VLDB} Endow.}, 11\penalty0 (12):\penalty0 1781--1794, 2018.
\newblock \doi{10.14778/3229863.3229867}.

\bibitem[Elouataoui et~al.(2022)Elouataoui, Alaoui, Mendili, and Gahi]{Elouataoui2022}
Widad Elouataoui, Imane~El Alaoui, Saida~El Mendili, and Youssef Gahi.
\newblock {An Advanced Big Data Quality Framework Based on Weighted Metrics}.
\newblock \emph{Big Data Cogn. Comput.}, 6\penalty0 (4):\penalty0 153, 2022.
\newblock \doi{10.3390/BDCC6040153}.

\bibitem[Heidari et~al.(2019)Heidari, McGrath, Ilyas, and Rekatsinas]{holodetect}
Alireza Heidari, Joshua McGrath, Ihab~F. Ilyas, and Theodoros Rekatsinas.
\newblock {HoloDetect: Few-Shot Learning for Error Detection}.
\newblock In \emph{{SIGMOD} Conference}, pages 829--846. {ACM}, 2019.
\newblock \doi{10.1145/3299869.3319888}.

\bibitem[Mahdavi et~al.(2019)Mahdavi, Abedjan, Fernandez, Madden, Ouzzani, Stonebraker, and Tang]{raha}
Mohammad Mahdavi, Ziawasch Abedjan, Raul~Castro Fernandez, Samuel Madden, Mourad Ouzzani, Michael Stonebraker, and Nan Tang.
\newblock {Raha: {A} Configuration-Free Error Detection System}.
\newblock In \emph{{SIGMOD} Conference}, pages 865--882. {ACM}, 2019.
\newblock \doi{10.1145/3299869.3324956}.

\bibitem[Shrestha et~al.(2023)Shrestha, Habibelahian, Termehchy, and Papotti]{Shrestha2023}
Rajesh Shrestha, Omeed Habibelahian, Arash Termehchy, and Paolo Papotti.
\newblock {Exploratory Training: When Annonators Learn About Data}.
\newblock \emph{Proc. {ACM} Manag. Data}, 1\penalty0 (2):\penalty0 135:1--135:25, 2023.
\newblock \doi{10.1145/3589280}.

\bibitem[Bors et~al.(2018)Bors, Gschwandtner, Kriglstein, Miksch, and Pohl]{Bors2018}
Christian Bors, Theresia Gschwandtner, Simone Kriglstein, Silvia Miksch, and Margit Pohl.
\newblock {Visual Interactive Creation, Customization, and Analysis of Data Quality Metrics}.
\newblock \emph{{ACM} J. Data Inf. Qual.}, 10\penalty0 (1):\penalty0 3:1--3:26, 2018.
\newblock \doi{10.1145/3190578}.

\bibitem[Restat et~al.(2023)Restat, Klettke, and St{\"{o}}rl]{CheDDaR}
Valerie Restat, Meike Klettke, and Uta St{\"{o}}rl.
\newblock {FAIR is not enough - A Metrics Framework to ensure Data Quality through Data Preparation}.
\newblock In \emph{{BTW}}, volume {P-331} of \emph{{LNI}}, pages 917--929. Gesellschaft f{\"{u}}r Informatik e.V., 2023.
\newblock \doi{10.18420/BTW2023-61}.

\bibitem[Diestelk{\"{a}}mper et~al.(2025)Diestelk{\"{a}}mper, Diestelk{\"{a}}mper, and Restat]{CheDDaR2025}
Indra Diestelk{\"{a}}mper, Ralf Diestelk{\"{a}}mper, and Valerie Restat.
\newblock {CheDDaR: Checking Data - Data Quality Report}.
\newblock In \emph{{BTW}}, volume {P-361} of \emph{{LNI}}, pages 1055--1067. Gesellschaft f{\"{u}}r Informatik e.V., 2025.
\newblock \doi{10.18420/BTW2025-70}.

\bibitem[Murta et~al.(2014)Murta, Braganholo, Chirigati, Koop, and Freire]{Murta2014}
Leonardo Murta, Vanessa Braganholo, Fernando Chirigati, David Koop, and Juliana Freire.
\newblock {noWorkflow: Capturing and Analyzing Provenance of Scripts}.
\newblock In \emph{{IPAW}}, volume 8628 of \emph{Lecture Notes in Computer Science}, pages 71--83. Springer, 2014.
\newblock \doi{10.1007/978-3-319-16462-5\_6}.

\bibitem[Chapman et~al.(2020)Chapman, Missier, Simonelli, and Torlone]{Chapman2020}
Adriane Chapman, Paolo Missier, Giulia Simonelli, and Riccardo Torlone.
\newblock Capturing and querying fine-grained provenance of preprocessing pipelines in data science.
\newblock \emph{Proc. {VLDB} Endow.}, 14\penalty0 (4):\penalty0 507--520, 2020.
\newblock \doi{10.14778/3436905.3436911}.

\bibitem[Grafberger et~al.(2022)Grafberger, Groth, Stoyanovich, and Schelter]{Grafberger2022}
Stefan Grafberger, Paul Groth, Julia Stoyanovich, and Sebastian Schelter.
\newblock {Data distribution debugging in machine learning pipelines}.
\newblock \emph{{VLDB} J.}, 31\penalty0 (5):\penalty0 1103--1126, 2022.
\newblock \doi{10.1007/S00778-021-00726-W}.

\bibitem[Biswas and Rajan(2021)]{Biswas2021}
Sumon Biswas and Hridesh Rajan.
\newblock {Fair preprocessing: towards understanding compositional fairness of data transformers in machine learning pipeline}.
\newblock In \emph{{ESEC/SIGSOFT} {FSE}}, pages 981--993. {ACM}, 2021.
\newblock \doi{10.1145/3468264.3468536}.

\bibitem[Zelaya(2019)]{Zelaya2019}
Carlos Vladimiro~Gonzalez Zelaya.
\newblock {Towards Explaining the Effects of Data Preprocessing on Machine Learning}.
\newblock In \emph{{ICDE}}, pages 2086--2090. {IEEE}, 2019.
\newblock \doi{10.1109/ICDE.2019.00245}.

\bibitem[Epperson et~al.(2024)Epperson, Gorantla, Moritz, and Perer]{Epperson2024}
Will Epperson, Vaishnavi Gorantla, Dominik Moritz, and Adam Perer.
\newblock {Dead or Alive: Continuous Data Profiling for Interactive Data Science}.
\newblock \emph{{IEEE} Trans. Vis. Comput. Graph.}, 30\penalty0 (1):\penalty0 197--207, 2024.
\newblock \doi{10.1109/TVCG.2023.3327367}.

\bibitem[Abedjan(2019)]{Abedjan2019}
Ziawasch Abedjan.
\newblock {Data Profiling}.
\newblock In \emph{Encyclopedia of Big Data Technologies}. Springer, 2019.
\newblock \doi{10.1007/978-3-319-63962-8\_8-1}.

\bibitem[Heine et~al.(2019)Heine, Kleiner, and Oelsner]{Heine2019}
Felix Heine, Carsten Kleiner, and Thomas Oelsner.
\newblock {Automated Detection and Monitoring of Advanced Data Quality Rules}.
\newblock In \emph{{DEXA} {(1)}}, volume 11706 of \emph{Lecture Notes in Computer Science}, pages 238--247. Springer, 2019.
\newblock \doi{10.1007/978-3-030-27615-7\_18}.

\bibitem[Tu et~al.(2023)Tu, He, Cui, Ge, Zhang, Han, Zhang, and Chaudhuri]{Tu2023}
Dezhan Tu, Yeye He, Weiwei Cui, Song Ge, Haidong Zhang, Shi Han, Dongmei Zhang, and Surajit Chaudhuri.
\newblock {Auto-Validate by-History: Auto-Program Data Quality Constraints to Validate Recurring Data Pipelines}.
\newblock In \emph{{KDD}}, pages 4991--5003. {ACM}, 2023.
\newblock \doi{10.1145/3580305.3599776}.

\bibitem[Shankar et~al.(2023)Shankar, Fawaz, Gyllstrom, and Parameswaran]{Shankar2023}
Shreya Shankar, Labib Fawaz, Karl Gyllstrom, and Aditya~G. Parameswaran.
\newblock {Automatic and Precise Data Validation for Machine Learning}.
\newblock In \emph{{CIKM}}, pages 2198--2207. {ACM}, 2023.
\newblock \doi{10.1145/3583780.3614786}.

\bibitem[Ehrlinger et~al.(2019)Ehrlinger, Haunschmid, Palazzini, and Lettner]{Ehrlinger2019}
Lisa Ehrlinger, Verena Haunschmid, Davide Palazzini, and Christian Lettner.
\newblock {A {DaQL} to Monitor Data Quality in Machine Learning Applications}.
\newblock In \emph{{DEXA} {(1)}}, volume 11706 of \emph{Lecture Notes in Computer Science}, pages 227--237. Springer, 2019.
\newblock \doi{10.1007/978-3-030-27615-7\_17}.

\bibitem[Lettner et~al.(2020)Lettner, Stumptner, Fragner, Rauchenzauner, and Ehrlinger]{Lettner2020}
Christian Lettner, Reinhard Stumptner, Werner Fragner, Franz Rauchenzauner, and Lisa Ehrlinger.
\newblock {{DaQL} 2.0: Measure Data Quality based on Entity Models}.
\newblock In \emph{{ISM}}, volume 180 of \emph{Procedia Computer Science}, pages 772--777. Elsevier, 2020.
\newblock \doi{10.1016/J.PROCS.2021.01.327}.

\bibitem[Narayanan et~al.(2024)Narayanan, S, and Zephan]{Narayanan2024}
Shammy Narayanan, Maheswari S, and Prisha Zephan.
\newblock {Real-Time Monitoring of Data Pipelines: Exploring and Experimentally Proving that the Continuous Monitoring in Data Pipelines Reduces Cost and Elevates Quality}.
\newblock \emph{ICST Transactions on Scalable Information Systems}, 2024.

\bibitem[Schuler et~al.(2023)Schuler, Singla, Vallat, White, Berman, and Kesselman]{DBLP:conf/eScience/SchulerSVWBK23}
Robert~E. Schuler, Jitin Singla, Brinda Vallat, Kate~L. White, Helen~M. Berman, and Carl Kesselman.
\newblock {Database Evolution, by Scientists, for Scientists: {A} Case Study}.
\newblock In \emph{e-Science}, pages 1--10. {IEEE}, 2023.
\newblock \doi{10.1109/E-SCIENCE58273.2023.10254872}.

\bibitem[Scherzinger and Sidortschuck(2020)]{DBLP:conf/er/ScherzingerS20}
Stefanie Scherzinger and Sebastian Sidortschuck.
\newblock {An Empirical Study on the Design and Evolution of {NoSQL} Database Schemas}.
\newblock In \emph{{ER}}, volume 12400 of \emph{Lecture Notes in Computer Science}, pages 441--455. Springer, 2020.
\newblock \doi{10.1007/978-3-030-62522-1\_33}.

\bibitem[Curino et~al.(2008)Curino, Moon, Tanca, and Zaniolo]{DBLP:conf/iceis/CurinoMTZ08}
Carlo Curino, Hyun~Jin Moon, Letizia Tanca, and Carlo Zaniolo.
\newblock {Schema Evolution in Wikipedia - Toward a Web Information System Benchmark}.
\newblock In \emph{{ICEIS} {(1)}}, pages 323--332, 2008.

\bibitem[Frei et~al.(2013)Frei, McWilliam, Derrick, Purvis, Tiwari, and Serugendo]{Frei2013SelfhealingAS}
Regina Frei, Richard McWilliam, Benjamin Derrick, Alan Purvis, Asutosh Tiwari, and Giovanna Di~Marzo Serugendo.
\newblock {Self-healing and self-repairing technologies}.
\newblock \emph{The International Journal of Advanced Manufacturing Technology}, 69:\penalty0 1033--1061, 2013.
\newblock URL \url{https://api.semanticscholar.org/CorpusID:39826532}.

\bibitem[Scherzinger et~al.(2021)Scherzinger, Mauerer, and Kondylakis]{DBLP:conf/icde/ScherzingerMK21}
Stefanie Scherzinger, Wolfgang Mauerer, and Haridimos Kondylakis.
\newblock {DeBinelle: Semantic Patches for Coupled Database-Application Evolution}.
\newblock In \emph{{ICDE}}, pages 2697--2700. {IEEE}, 2021.
\newblock \doi{10.1109/ICDE51399.2021.00307}.

\bibitem[Schuler and Kesselman(2022)]{DBLP:conf/eScience/SchulerK22}
Robert~E. Schuler and Carl Kesselman.
\newblock {Managing Database-Application Co-Evolution in a Scientific Data Ecosystem}.
\newblock In \emph{e-Science}, pages 214--224. {IEEE}, 2022.
\newblock \doi{10.1109/ESCIENCE55777.2022.00035}.

\bibitem[Calinescu et~al.(2020)Calinescu, Mirandola, Perez{-}Palacin, and Weyns]{uncertainty1}
Radu Calinescu, Raffaela Mirandola, Diego Perez{-}Palacin, and Danny Weyns.
\newblock {Understanding Uncertainty in Self-adaptive Systems}.
\newblock In \emph{{ACSOS}}, pages 242--251. {IEEE}, 2020.
\newblock \doi{10.1109/ACSOS49614.2020.00047}.

\bibitem[Mahdavi{-}Hezavehi et~al.(2021)Mahdavi{-}Hezavehi, Weyns, Avgeriou, Calinescu, Mirandola, and Perez{-}Palacin]{uncertainty2}
Sara Mahdavi{-}Hezavehi, Danny Weyns, Paris Avgeriou, Radu Calinescu, Raffaela Mirandola, and Diego Perez{-}Palacin.
\newblock {Uncertainty in Self-Adaptive Systems: {A} Research Community Perspective}.
\newblock \emph{CoRR}, abs/2103.02717, 2021.
\newblock URL \url{https://arxiv.org/abs/2103.02717}.

\bibitem[Klettke et~al.(2017)Klettke, Awolin, St{\"{o}}rl, M{\"{u}}ller, and Scherzinger]{DBLP:conf/bigdataconf/KlettkeAS0S17}
Meike Klettke, Hannes Awolin, Uta St{\"{o}}rl, Daniel M{\"{u}}ller, and Stefanie Scherzinger.
\newblock {Uncovering the Evolution History of Data Lakes}.
\newblock In \emph{{IEEE} BigData}, pages 2462--2471. {IEEE} Computer Society, 2017.
\newblock \doi{10.1109/BIGDATA.2017.8258204}.

\bibitem[Spoth et~al.(2021)Spoth, Kennedy, Lu, Hammerschmidt, and Liu]{DBLP:conf/sigmod/SpothKLHL21}
William Spoth, Oliver Kennedy, Ying Lu, Beda~Christoph Hammerschmidt, and Zhen~Hua Liu.
\newblock {Reducing Ambiguity in Json Schema Discovery}.
\newblock In \emph{{SIGMOD} Conference}, pages 1732--1744. {ACM}, 2021.
\newblock \doi{10.1145/3448016.3452801}.

\bibitem[Fu and Chen(2024)]{fu2024compoundschemaregistry}
Silvery~D. Fu and Xuewei Chen.
\newblock {Compound Schema Registry}.
\newblock \emph{CoRR}, abs/2406.11227, 2024.
\newblock \doi{10.48550/ARXIV.2406.11227}.

\bibitem[Shevtsov et~al.(2018)Shevtsov, Berekmeri, Weyns, and Maggio]{DBLP:journals/tse/ShevtsovBWM18}
Stepan Shevtsov, Mihaly Berekmeri, Danny Weyns, and Martina Maggio.
\newblock {Control-Theoretical Software Adaptation: {A} Systematic Literature Review}.
\newblock \emph{{IEEE} Trans. Software Eng.}, 44\penalty0 (8):\penalty0 784--810, 2018.
\newblock \doi{10.1109/TSE.2017.2704579}.

\bibitem[Ruiz(2024)]{DBLP:conf/ease/Ruiz24}
Fernando~Vallecillos Ruiz.
\newblock {Agent-Driven Automatic Software Improvement}.
\newblock In \emph{{EASE}}, pages 470--475. {ACM}, 2024.
\newblock \doi{10.1145/3661167.3661171}.

\bibitem[Yang et~al.(2013)Yang, Lu, Tao, Ma, Xing, and Song]{DBLP:journals/jcst/YangLTMXS13}
Qiliang Yang, Jian Lu, XianPing Tao, Xiaoxing Ma, Jianchun Xing, and Wei Song.
\newblock {Fuzzy Self-Adaptation of Mission-Critical Software Under Uncertainty}.
\newblock \emph{J. Comput. Sci. Technol.}, 28\penalty0 (1):\penalty0 165--187, 2013.
\newblock \doi{10.1007/S11390-013-1321-9}.

\bibitem[Kramer(2023)]{DBLP:conf/gvd/Kramer23}
Kevin Kramer.
\newblock {Towards Evolution Capabilities in Data Pipelines}.
\newblock In \emph{GvDB}, volume 3714 of \emph{{CEUR} Workshop Proceedings}. CEUR-WS.org, 2023.
\newblock URL \url{https://ceur-ws.org/Vol-3714/paper7.pdf}.

\bibitem[Krishnan et~al.(2016)Krishnan, Haas, Franklin, and Wu]{Krishnan2016}
Sanjay Krishnan, Daniel Haas, Michael~J. Franklin, and Eugene Wu.
\newblock {Towards reliable interactive data cleaning: a user survey and recommendations}.
\newblock In \emph{HILDA@SIGMOD}, page~9. {ACM}, 2016.
\newblock \doi{10.1145/2939502.2939511}.

\bibitem[Polyzotis and Zaharia(2021)]{Zaharia2021}
Neoklis Polyzotis and Matei Zaharia.
\newblock {What can Data-Centric {AI} Learn from Data and {ML} Engineering?}
\newblock \emph{CoRR}, abs/2112.06439, 2021.
\newblock URL \url{https://arxiv.org/abs/2112.06439}.

\bibitem[Johnson et~al.(2024)Johnson, Simon, and Aliferis]{Johnson2024}
Steven~G. Johnson, Gyorgy Simon, and Constantin Aliferis.
\newblock \emph{Data Preparation, Transforms, Quality, and Management}, pages 377--413.
\newblock Springer International Publishing, 2024.
\newblock \doi{10.1007/978-3-031-39355-6_8}.

\bibitem[Restat et~al.(2024)Restat, Klettke, and St{\"{o}}rl]{dataplat}
Valerie Restat, Meike Klettke, and Uta St{\"{o}}rl.
\newblock {Towards an End-to-End Data Quality Optimizer}.
\newblock In \emph{{ICDEW}}, pages 262--266. {IEEE}, 2024.
\newblock \doi{10.1109/ICDEW61823.2024.00039}.

\bibitem[Tsai and Hu(2022)]{Tsai22}
Chih{-}Fong Tsai and Ya{-}Han Hu.
\newblock {Empirical comparison of supervised learning techniques for missing value imputation}.
\newblock \emph{Knowl. Inf. Syst.}, 64\penalty0 (4):\penalty0 1047--1075, 2022.
\newblock \doi{10.1007/S10115-022-01661-0}.

\bibitem[Hasan et~al.(2021)Hasan, Alam, Roy, Dutta, Jawad, and Das]{Hasan2021}
Md.~Kamrul Hasan, Md.~Ashraful Alam, Shidhartho Roy, Aishwariya Dutta, Md.~Tasnim Jawad, and Sunanda Das.
\newblock {Missing value imputation affects the performance of machine learning: A review and analysis of the literature (2010–2021)}.
\newblock \emph{Informatics in Medicine Unlocked}, 27:\penalty0 100799, 2021.

\bibitem[Restat and St{\"{o}}rl(2025)]{ALPINE}
Valerie Restat and Uta St{\"{o}}rl.
\newblock {{ALPINE:} Abstract Language for Pipeline Integration and Execution}.
\newblock In \emph{{BTW} Workshops}, volume {P-363} of \emph{{LNI}}, pages 207--217. Gesellschaft f{\"{u}}r Informatik e.V., 2025.
\newblock \doi{10.18420/BTW2025-125}.

\bibitem[Strasser(2024)]{gvdb_sebastian}
Sebastian Strasser.
\newblock {Towards machine learning-aware data validation}.
\newblock In \emph{GvDB}, volume 3710 of \emph{{CEUR} Workshop Proceedings}, pages 29--34. CEUR-WS.org, 2024.
\newblock URL \url{https://ceur-ws.org/Vol-3710/paper5.pdf}.

\bibitem[Strasser and Klettke(2024)]{hilda}
Sebastian Strasser and Meike Klettke.
\newblock {Transparent Data Preprocessing for Machine Learning}.
\newblock In \emph{HILDA@SIGMOD}, pages 1--6. {ACM}, 2024.
\newblock \doi{10.1145/3665939.3665960}.

\bibitem[Breunig et~al.(2000)Breunig, Kriegel, Ng, and Sander]{Breunig2000}
Markus~M. Breunig, Hans{-}Peter Kriegel, Raymond~T. Ng, and J{\"{o}}rg Sander.
\newblock {LOF: Identifying Density-Based Local Outliers}.
\newblock In \emph{{SIGMOD} Conference}, pages 93--104. {ACM}, 2000.
\newblock \doi{10.1145/342009.335388}.

\bibitem[Kramer et~al.(2025)Kramer, Restat, and St{\"{o}}rl]{robustness}
Kevin~M. Kramer, Valerie Restat, and Uta St{\"{o}}rl.
\newblock Evolving gracefully: Building robust and self-adaptive data cleaning pipelines for schema evolution and uncertainty.
\newblock In \emph{{VLDB} Workshops}. VLDB.org, 2025.
\newblock Accepted for publication.

\bibitem[Kephart and Chess(2003)]{MAPE-K}
Jeffrey~O. Kephart and David~M. Chess.
\newblock {The Vision of Autonomic Computing}.
\newblock \emph{Computer}, 36\penalty0 (1):\penalty0 41--50, 2003.
\newblock \doi{10.1109/MC.2003.1160055}.

\bibitem[Zou(2020)]{Zou2020ASO}
Xiaohan Zou.
\newblock {A Survey on Application of Knowledge Graph}.
\newblock \emph{Journal of Physics: Conference Series}, 1487, 2020.
\newblock URL \url{https://api.semanticscholar.org/CorpusID:216342059}.

\bibitem[Mior(2024)]{mior2024largelanguagemodelsjson}
Michael~J. Mior.
\newblock {Large Language Models for {JSON} Schema Discovery}.
\newblock \emph{CoRR}, abs/2407.03286, 2024.
\newblock \doi{10.48550/ARXIV.2407.03286}.

\bibitem[Glake et~al.(2021)Glake, Kiehn, Schmidt, and Ritter]{DBLP:conf/summersoc/GlakeKSR21}
Daniel Glake, Felix Kiehn, Mareike Schmidt, and Norbert Ritter.
\newblock {Towards Taming the Adaptivity Problem - Formalizing Poly-/MultiStore Topology Descriptions}.
\newblock In \emph{SummerSOC}, volume 1429 of \emph{Communications in Computer and Information Science}, pages 83--99. Springer, 2021.
\newblock \doi{10.1007/978-3-030-87568-8\_5}.

\bibitem[Klettke et~al.(2015)Klettke, St{\"{o}}rl, and Scherzinger]{DBLP:conf/btw/KlettkeSS15}
Meike Klettke, Uta St{\"{o}}rl, and Stefanie Scherzinger.
\newblock {Schema Extraction and Structural Outlier Detection for JSON-based NoSQL Data Stores}.
\newblock In \emph{{BTW}}, volume {P-241} of \emph{{LNI}}, pages 425--444. {GI}, 2015.
\newblock URL \url{https://dl.gi.de/handle/20.500.12116/2420}.

\end{thebibliography}

\end{document}